\documentclass[10pt,twocolumn,aps,pra]{revtex4-2}

\usepackage{graphicx}
\usepackage{dcolumn}
\usepackage{bm}
\usepackage[dvipsnames]{xcolor}
\usepackage[utf8]{inputenc}
\usepackage{amsmath, amsfonts, amsthm, bbm}
\usepackage{amsthm}
\usepackage{mathtools}
\usepackage{amssymb}
\usepackage{pifont}
\usepackage{bm}
\usepackage{graphicx} 
\usepackage{bmpsize}
\usepackage{float}
\usepackage{braket}

\definecolor{brickred}{rgb}{0.8, 0.25, 0.33}
\newcommand\myshade{85}
\colorlet{mylinkcolor}{BrickRed}
\colorlet{mycitecolor}{NavyBlue}
\colorlet{myurlcolor}{Aquamarine}

\usepackage{hyperref}
\hypersetup{
  linkcolor  = mylinkcolor!\myshade!black,
  citecolor  = mycitecolor!\myshade!black,
  urlcolor   = myurlcolor!\myshade!black,
  colorlinks = true,
}

\def\br#1{\left[ #1 \right]}
\def\brpos#1{\left[ #1 \right]_+}
\def\pr#1{\left( #1\right)}

\def\ang#1{\left\langle #1\right\rangle}
\def\abs#1{\left| #1 \right|}
\newcommand{\iu}{\mathrm{i}\mkern1mu}

\DeclareMathOperator{\Tr}{Tr}
\DeclareMathOperator*{\argmax}{arg\,max}

\let\emptyset\varnothing

\makeatletter
\newcommand{\vast}{\bBigg@{3}}
\newcommand{\Vast}{\bBigg@{4}}
\makeatother

\makeatletter
\newsavebox{\@brx}
\newcommand{\llangle}[1][]{\savebox{\@brx}{\(\m@th{#1\langle}\)}%
  \mathopen{\copy\@brx\mkern2mu\kern-0.9\wd\@brx\usebox{\@brx}}}
\newcommand{\rrangle}[1][]{\savebox{\@brx}{\(\m@th{#1\rangle}\)}%
  \mathclose{\copy\@brx\mkern2mu\kern-0.9\wd\@brx\usebox{\@brx}}}
\makeatother

\begin{document}

\preprint{APS/123-QED}

\title{Explosive neural networks via higher-order interactions in curved statistical manifolds}

\renewcommand{\andname}{\ignorespaces}

\author{Miguel Aguilera}
\email{maguilera@bcamath.org}
\affiliation{
BCAM -- Basque Center for Applied Mathematics, Bilbao, Spain\\
IKERBASQUE, Basque Foundation for Science, Bilbao, Spain
}
\author{Pablo A. Morales}
\affiliation{Research Division, Araya Inc., Tokyo, Japan}
\affiliation{Centre for Complexity Science, Imperial College London, London, UK
}
\author{Fernando E. Rosas}
\affiliation{Sussex AI and Sussex Centre for Consciousness Science, Department of Informatics, University of Sussex, Brighton, UK\\
Department of Brain Sciences and Centre for Complexity Science, Imperial College London, London, UK\\
Center for Eudaimonia and Human Flourishing, University of Oxford, Oxford, UK\\
Principles of Intelligent Behavior in Biological and Social Systems (PIBBSS), Prague, 
Czech Republic
}
\author{Hideaki Shimazaki}
\affiliation{Graduate School of Informatics, Kyoto University, Kyoto, Japan\\
Center for Human Nature, Artificial Intelligence, and Neuroscience (CHAIN),  Hokkaido University, Sapporo, Japan
}

\begin{abstract}
\vspace{1mm}
\section*{Abstract}
\noindent
Higher-order interactions underlie complex phenomena in systems such as biological and artificial neural networks, but their study is challenging due to the scarcity of tractable models. By leveraging a generalisation of the maximum entropy principle, we introduce \emph{curved neural networks} as a class of models with a limited number of parameters that are particularly well-suited for studying higher-order phenomena. Through exact mean-field descriptions, we show that these curved neural networks implement a self-regulating annealing process that can accelerate memory retrieval, leading to explosive order-disorder phase transitions with multi-stability and hysteresis effects. Moreover, by analytically exploring their memory-retrieval capacity using the replica trick, we demonstrate that these networks can enhance memory capacity and robustness of retrieval over classical associative-memory networks. Overall, the proposed framework provides parsimonious models amenable to analytical study, revealing higher-order phenomena in complex networks.
\end{abstract}

\maketitle

\section*{Introduction}
Complex physical, biological, and social systems often exhibit higher-order interdependencies that cannot be reduced to pairwise interactions between their components~\cite{lambiotte2019networks,battiston2021physics}.
Recent studies suggest that higher-order organisation is not the exception but the norm, providing various mechanisms for its emergence~\cite{amari2003synchronous,kuehn2021universal,shomali2023uncovering,thibeault2024low}.
Modelling studies have revealed that higher-order interactions (HOIs) underlie collective activities such as bistability, hysteresis, and `explosive' phase transitions associated with abrupt discontinuities in order parameters~\cite{angst2012explosive,d2019explosive,iacopini2019simplicial,millan2020explosive,kuehn2021universal,landry2020effect}.

HOIs are particularly important for the functioning of biological and artificial neural systems. For instance, they shape the collective activity of biological neurons ~\cite{montani2009impact,tkavcik2014searching}, being directly responsible for their 
inherent sparsity~\cite{ohiorhenuan2010sparse, tkavcik2014searching, shimazaki2015simultaneous, shomali2023uncovering} and possibly underlying critical dynamics~\cite{tkavcik2013simplest, tkavcik2015thermodynamics}. HOIs have also been shown to enhance the computational capacity of artificial recurrent neural networks~\cite{burns2022simplicial, bybee2023efficient}. More specifically, `dense associative memories' with extended memory capacity~\cite{krotov2016dense,demircigil2017model,agliari2023dense,lucibello2024exponential} are realised by specific non-linear activation functions, which effectively incorporate HOIs. These non-linear functions are related to attention mechanisms of transformer neural networks~\cite{krotov2023new} and the energy landscape of diffusion models~\cite{ambrogioni2024search,ambrogioni2023statistical}, leading to the conjecture that HOIs underlie the success of these state-of-the-art deep learning models. 

Despite their importance, existent studies of HOIs face significant computational challenges. Analytically tractable models that incorporate HOIs typically limit interactions to a single order (e.g., p-spin models \cite{bovier2001spin,agliari2022nonlinear,agliari2023dense}). Otherwise, attempting to represent diverse HOIs exhaustively results in a combinatorial explosion~\cite{amari2001information}.
This issue is pervasive, restricting investigations of high-order interaction models --- such as contagion \cite{iacopini2019simplicial}, Ising \cite{bybee2023efficient}, or Kuramoto \cite{skardal2020higher} models --- to highly homogeneous scenarios~\cite{amari2003synchronous,tkavcik2013simplest} or to models of relatively low-order~\cite{ganmor2011sparse,iacopini2019simplicial,landry2020effect}. 
While attempts have been made to model all orders of HOIs and perform theoretical analyses~\cite{krotov2016dense,demircigil2017model,agliari2023dense,lucibello2024exponential,barra2018new,agliari2019relativistic,agliari2020generalized,rodriguez2023alternating,santos2024hopfield,hoover2024dense}, it is currently unclear how to construct parsimonious models to address the diverse effects of HOIs in a principled manner.

To address this challenge, here we employ an extension of the maximum entropy principle to capture HOIs through the deformation of the space of statistical models. When applied to neural networks, our approach generalises classical neural network models to yield a family of \emph{curved neural networks} that effectively incorporate HOIs of all orders. The resulting models have rich connections with the literature on the statistical physics of neural networks~\cite{bovier2001spin,agliari2020generalized,agliari2023dense,demircigil2017model}. These features enable the exploration of various aspects of HOIs using techniques including mean-field approximations, quenched disorder analyses, and path integrals.

Our analyses reveal how relatively simple curved neural networks exhibit some of the hallmark characteristics of higher-order phenomena, such as explosive phase transitions, arising both in mean-field models and in more complex transitions to spin-glass states. These phenomena are driven by a self-regulated annealing process, which accelerates memory retrieval through positive feedback between energy and an `effective' temperature --- a perspective that can also explain memory-retrieval dynamics in other modern artificial networks. Furthermore, we show --- both analytically and experimentally --- that this mechanism can lead to an increase in the memory capacity or robustness of memory retrieval in these neural networks.
Overall, the core contributions of this work are (i) the development of a parsimonious neural network model based on the maximum entropy principle that captures interactions of all orders, (ii) the discovery of a self-regulated annealing mechanism that can drive explosive phase transitions, and (iii) the demonstration of enhanced memory capacity resulting from this mechanism.

\section*{Results}

\subsection*{High-order interactions in curved manifolds}


The maximum entropy principle (MEP) is a general modelling framework based on the principle of adopting the model with maximal entropy compatible with a given set of observations, under the rationale that one should not assume any structure beyond what is specified by the assumptions or features selected from the data~\cite{jaynes2003probability,cofre2019comparison}. The traditional formulation of the MEP is based on Shannon's entropy~\cite{jaynes1957information}, and the resulting models correspond to Boltzmann distributions of the form $p(\bm x) = \exp\pr{\sum_a \theta_a f_a(\bm x) - \varphi}$, where $\bm x=(x_1,\dots,x_n)$, $\varphi$ is a normalising potential and $\theta_a$ are parameters constraining the average value of observables $\ang{f_a(\bm x)}$. While observables are often set to low orders (e.g. $f_i(\bm x)=x_i$, $f_{ij}(\bm x)=x_ix_j$, corresponding to first and second order statistics), higher-order interdependencies can be included by considering observables of the type $f_{\bm I}(\bm x) = \prod_{i\in\bm I} x_i$, where $\bm I$ is a set of indices of order $k= |\bm I|$. Unfortunately, an exhaustive description of interactions up to order $k\gg 1$ becomes unfeasible in practice due to an exponential number of terms (for more details on the MEP, see Supplementary Note~1).

The MEP can be expanded to include other entropy functionals such as Tsallis'~\cite{tsallis1998role} and Rényi's~\cite{morales2021generalization}. Concretely, maximising the Rényi entropy (with the scaling parameter $\gamma \geq -1)$ ~\cite{valverde2019case}
\begin{equation}
    H_\gamma(p) = - \frac{1}{\gamma} \ln \sum_{\bm x} p(\bm x)^{1+\gamma}
\end{equation}
while constraining $\ang{f_a(\bm x)}$ (i.e., the expectation of features by $p(\bm x)$) results in models of the form (see Supplementary Note~1):
\begin{equation}
    p_{\gamma}(\bm x) = \exp(-\varphi_{\gamma}) \bigg[1 + \gamma  \beta\sum_a \theta_a f_a(\bm x)\bigg]_+^{1/\gamma} ,
\label{eq:deformed_exp}
\end{equation}
where $\varphi_{\gamma}$ is a normalising constant given by
\begin{equation}
    \varphi_{\gamma} = \ln \sum_{\bm x} \bigg[1 + \gamma  \beta\sum_a \theta_a f_a(\bm x)\bigg]_+^{1/\gamma}. 
\end{equation}
Above, the square bracket operator sets negative values to zero, $\brpos{x}=\max\{0,x\}$. 
We refer to distributions following \eqref{eq:deformed_exp} as the \emph{deformed exponential family} distributions, which maximises both Rényi and Tsallis entropies~\cite{umarov2008aq,wong2022tsallis}. When $\gamma\rightarrow 0$, Rényi's entropy tends to Shannon's and \eqref{eq:deformed_exp} to the standard exponential family~\cite{morales2021generalization}.

\begin{figure}
    \centering
    \includegraphics[width=8cm]{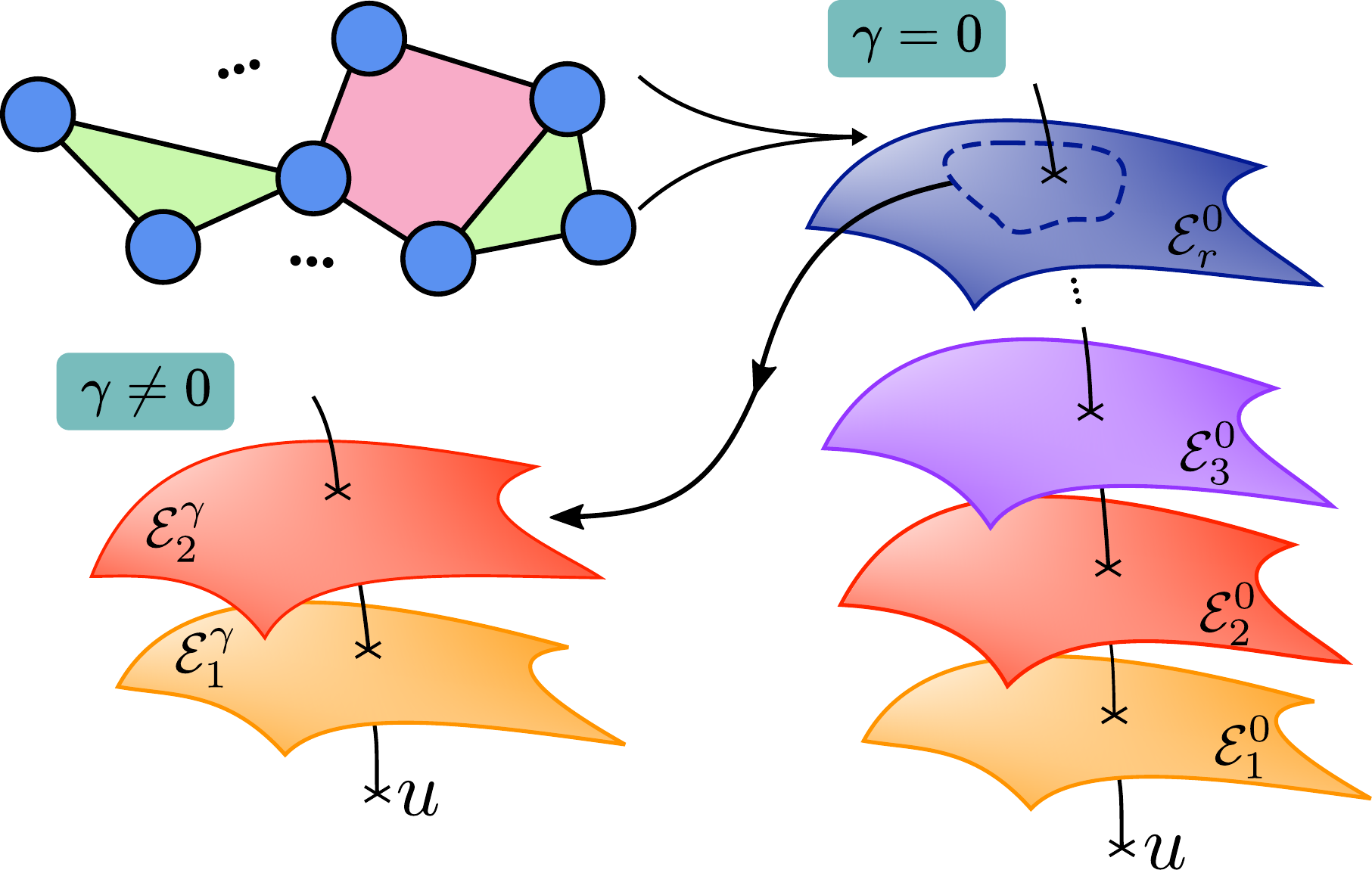}
    \caption{\textbf{Higher-order decomposition resulting from the foliation of a statistical manifold.} Illustration of a family of standard MEP models (right) and its deformed counterpart (bottom left). The space of MEP distributions with constraints of different orders constitute nested sub-manifolds~\cite{amari2001information}, 
    giving rise to a hierarchy of sub-families of models of the form $\mathcal{E}_k^\gamma = 
    \{ p_\gamma^{(k)}(\bm x) = e^{-\varphi_{\gamma}}  \big[1 - \gamma \beta E_k(\bm x) \big]_+^{1/\gamma} 
    \}$ such that $\mathcal{E}_1^\gamma\subset\mathcal{E}_2^\gamma\subset\dots\subset\mathcal{E}_n^\gamma$ \cite{morales2021generalization}.
    The foliation depends on the curvature $\gamma$, and in general $\mathcal{E}_k^\gamma \neq \mathcal{E}_k^0$ but rather $\mathcal{E}_k^\gamma \cap \mathcal{E}_r^0 \neq \emptyset$ for $k < r$.
    For small values of  $|\gamma|$, it is possible to neglect higher-order terms in \eqref{eq:deformed_exp_expansion}, and therefore certain subsets of $\mathcal{E}_k^\gamma$ effectively approximate $\mathcal{E}_r^0$.
    } 
    \label{fig:order_mixing}
\end{figure}

A fundamental insight explored in this study is that higher-order interdependencies can be efficiently captured by deformed exponential family distributions \cite{guisande2024renyi,jauregui2018characterization}. Starting from a standard Shannon's MEP model with low-order interactions, it can be shown that varying $\gamma$ in \eqref{eq:deformed_exp} results in a deformation of the statistical manifold which, in turn, enhances the capability of $p_\gamma(\bm x)$ to account for higher-order interdependencies. In effect, the consequence of deformation can be investigated by rewriting \eqref{eq:deformed_exp} via Taylor expansion of the exponent
\begin{equation}
    p_\gamma(\bm x) = \exp\bigg( \sum_{k=1}^\infty \frac{-1}{k\gamma} \Big( -\gamma \beta \sum_a \theta_a f_a(\bm x)\Big)^k -\varphi_{\gamma}\bigg),
    \label{eq:deformed_exp_expansion}
\end{equation}
which is valid for the case $1+\gamma \sum_a \theta_a f_a(\bm x)>0$, and otherwise $p_\gamma(\bm x)=0$. 
This shows that the deformed manifold contains interactions of all orders even if $f_a(\bm x)$ is restricted to lower orders while establishing a specific dependency structure across the orders, thereby avoiding a combinatorial explosion of the number of required parameters. 
The deformation resulting from the maximisation of a non-Shannon entropy has been shown to reflect a curvature of the space of possible models in information geometry~\cite{wong2018logarithmic,vigelis2019properties,morales2021generalization,wong2022tsallis}. This leads to a particular \textit{foliation} of the space of possible models~\cite{amari2016information} (an `onion-like' manifold structure, Fig.~\ref{fig:order_mixing}), which has properties that allow to re-derive the MEP from fundamental geometric properties --- for technical details, see Supplementary Note~1.

\subsection*{Curved neural networks}

Several well-known neural network models adhere to the MEP, such as Ising-like models \cite{roudi2015multi} and Boltzmann machines \cite{montufar2018restricted}. Interestingly, these models can encode patterns in their weights in the form of `associative memories' as in Nakano-Amari-Hopfield networks~\cite{nakano1972associatron,amari1972learning,hopfield1982neural}, being amenable for investigations using tools from equilibrium and nonequilibrium statistical physics literature~\cite{amit1989modeling, coolen2005theory, coolen2001statistical, coolen2001statisticalII}.
Following the principles laid down in the previous section, we now introduce a family of recurrent neural networks that we call \emph{curved neural networks}.

For this purpose, let us consider $N$ binary variables $x_1,\dots,x_N$ taking values in $\{1,-1\}$ following a joint probability distribution
\begin{equation}
    p_\gamma(\bm x) = \exp\pr{-\varphi_{\gamma}}\big[1-\gamma \beta E(\bm x)\big]_+^{1/\gamma},
    \label{eq:deformed-Hopfield-net}
\end{equation}
where $\varphi_{\gamma}$ is a normalising constant. 
Above, we call $E(\bm x)$ and $\beta$ the (stochastic) \emph{energy function} (i.e., Hamiltonian) and the \emph{inverse temperature} due to their similarity with the Gibbs distribution in statistical physics when $\gamma\to 0$. Note that, unlike exponential families, these models do not exhibit energy invariance under constant shifts. However, as demonstrated in Ref.~\cite{tsallis1998role}, deformed exponential models can be related to energy-invariant models by rescaling their temperature, which can be seen as maximising entropy with respect to escort statistics rather than the original natural statistics. 

Neural network models are typically defined by considering $p_\gamma(\bm x)$ as defined in \eqref{eq:deformed-Hopfield-net} with an energy function of the form
\begin{equation}
    E(\bm x)= - \sum_{i=1}^N H_i x_i  - \frac{1}{N} \sum_{i<j} J_{ij} x_i x_j,
\end{equation}
where $J_{ij}$ is the coupling strength between neurons $x_i$ and $x_j$, and $H_i$ are bias terms. 
In the limit $\gamma\rightarrow 0$, $p_0(\bm x)$ recovers the Ising model.
Emulating  classical associative memories, the weights $J_{ij}$ can be made to encode a collection of $M$ neural patterns $\bm\xi^a = \{\xi_1^a, \dots \xi_N^a\}$,  $\xi_1^a=\pm 1$ and $a=1,\dots,M$ by using the well-known Hebbian rule \cite{hopfield1982neural,amit1989modeling}
\begin{equation}
    J_{ij} =J \sum_{a=1}^M \xi_i^a \xi_j^a,
    \label{eq:weight-encoding}
\end{equation}
where $J$ is a scaling parameter. 

Before proceeding with our main analysis, one can gain insights into the effect of the curvature 
$\gamma$ from the dynamics of a recurrent neural network that behaves as a sampler of the equilibrium distribution described by \eqref{eq:deformed-Hopfield-net}. For this, we adapt the classic Glauber dynamics to curved neural networks (see Supplementary Note~2) to obtain
\begin{equation}
    p(x_i | \bm{x}_{\backslash i}) =
    \pr{1 +\brpos{1 - \gamma \beta'(\bm x) \Delta E(\bm x)}^{1/\gamma}}^{-1}, 
    \label{eq:Glauber}
\end{equation}
where $\bm{x}_{\backslash i}$ denotes the state of all neurons except $x_i$, $\Delta E(\bm x)=2 x_i (H_i + \frac{1}{N}\sum_j J_{ij} x_j)$ is the energy difference associated with detailed balance, and $\beta'(\bm x)$ is an effective inverse temperature given by
\begin{equation}
    \beta'(\bm x) = \frac{\beta}{\brpos{1-\gamma \beta E(\bm x)}}.
    \label{eq:beta_Glauber}
\end{equation}
Again, $\gamma \to 0$ recovers the classic Glauber dynamics and $\beta'(\bm x)=\beta$. 
Thus, the curvature affects the dynamics through the deformed nonlinear activation function \eqref{eq:Glauber} and the state-dependent effective temperature $\beta'(\bm x)$ \eqref{eq:beta_Glauber}, with higher $\beta'(\bm x)$ inducing lower degrees of randomness in the transitions. The effect of $E(\bm x)$ on $\beta'(\bm x)$ depends then on the sign of $\gamma$. A negative $\gamma$ increases $\beta'(\bm{x})$ during relaxation, reducing the stochasticity of the dynamics and accelerating convergence to a low-energy state. This, in turn, raises $\beta'$, creating a positive feedback loop between energy and effective temperature. 
The effect is similar to simulated annealing, but the coupling of the energy and effective inverse temperature lets the annealing scheduling self-regulate to accelerate convergence. In contrast, positive $\gamma$ decelerates the dynamics through negative feedback. Such accelerating or decelerating dynamics underlie non-trivial complex collective behaviours of the curved neural networks, which will be examined in the subsequent sections.

\subsection*{Mean-field behaviour of curved associative-memory networks}


As with regular associative memories \cite{coolen2001statistical}, one can solve the behaviour of curved associative-memory networks through mean-field methods in the thermodynamic limit $N\to\infty$ (Supplementary Note~3). Here the energy is extensive, meaning that it scales with the system's size $N$. To ensure the deformation parameter remains independent of system properties such as size or temperature, we scale it as follows:
\begin{equation}
    \gamma = \frac{\gamma'}{N\beta}.
\end{equation}

Under this condition, we calculate the normalising potential $\varphi_{\gamma}$ by introducing a delta integral and calculating a saddle-node solution, resulting in a set of order parameters $\bm m = \{m_1,\ldots,m_M\}$, $m_a=\frac{1}{N}\sum_i \xi_i^a \ang{x_i}$ in the limit of size $N\to\infty$. This calculation assumes $1-\gamma \beta E(\bm x)>0$ so that $\brpos{\,}$ operators can be omitted and $\varphi_{\gamma}$ is differentiable. The solution results in (for $H_i=0$):
\begin{align}
    \varphi_{\gamma} &=  N\frac{\beta }{\gamma'} \ln \frac{\beta'}{\beta} - \sum_{a=1}^M  \beta' N J m_a^2 
    \nonumber\\& \,\quad + \sum_{i=1}^N \ln \bigg(2 \cosh\bigg(\beta' J \sum_{a=1}^M \xi_i^a  m_a\bigg)\bigg), \label{eq:neg_norm_potential}
\end{align}
where $\beta'$ is given by
\begin{equation}
    \beta' = \frac{ \beta}{1+\gamma' \frac{1}{2} J \sum_a  m_a^2},\label{eq:mf-Hopfield-beta}
\end{equation}
and the values of the mean-field variables $m_a$ are found from the following self-consistent equations:
\begin{equation}
    m_a =\sum_{i=1}^N  \frac{\xi_i^a}{N} \tanh \bigg(\beta'J\sum_{b=1}^M \xi_i^b  m_b\bigg).
    \label{eq:mf-Hopfield}
\end{equation}

Similarly, using a generating functional approach \cite{coolen2001statisticalII}, we use the Glauber rule in \eqref{eq:Glauber} to derive a dynamical mean-field given by path integral methods (see Supplementary Note~4).
This yields
\begin{equation}
    \dot m_{a}= -m_{a}
    + \sum_{i=1}^N \frac{\xi_i^a}{N}  \tanh\bigg(\beta'J\sum_{b=1}^M \xi_i^b m_{b}\bigg),
    \label{eq:dynamical-mean-field}
\end{equation}
where $\beta'$ is defined as in \eqref{eq:mf-Hopfield-beta} for each $\bm m$. Note that in large systems, we recover the classical nonlinear activation function, and the deformation affects the dynamics only through the effective temperature $\beta'$.

\subsection*{Explosive phase transitions}

\begin{figure*}[!t]
    \centering
    \includegraphics[width=17cm]{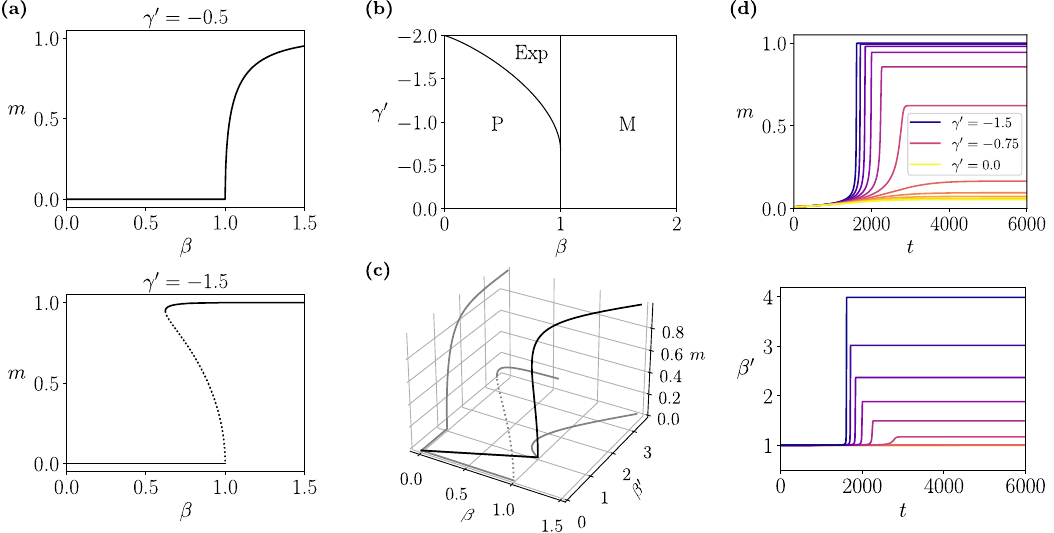}
    \caption{ \textbf{Explosive phase transitions in curved neural networks. }
    \textbf{(a)} Phase transitions of the curved neural network with one associative memory, for $J=1$ and values of $\gamma'=-0.5$ (top, displaying a second-order phase transition) and $\gamma'=-1.5$ (bottom, displaying an explosive phase transition). Solid lines represent the stable fixed points, and dotted lines correspond to unstable fixed points.
    \textbf{(b)} Phase diagram of the system. The areas indicated by P and M refer to the usual paramagnetic (disordered) and magnetic (ordered) phases, respectively. The area indicated by Exp represents a phase where ordered and disordered states coexist in an explosive phase transition characterised by a hysteresis loop. \textbf{(c)} Solutions of (\ref{eq:mf_one-pattern}-\ref{eq:beta_one-pattern}) for $\beta',m,\beta$ (black line) for $\gamma'=-1.2$, and projections to the plane $m=0$, $\beta=0$ and $\beta'=0 $, obtaining respectively the relation between $\beta,\beta'$ and solutions of the flat and the deformed models respectively (grey lines).
     \textbf{(d)}  Mean-field dynamics of the single-pattern neural network for $\beta=1.001$ (near criticality from the ordered phase) for some values of $\gamma'$ in $[-1.5,0]$. For large negative $\gamma'$ the dynamics `explodes', with $m$ (top) and $\beta'$ (bottom) converging abruptly.} 
    \label{fig:explosive-phase-transitions}
\end{figure*}

To illustrate these findings, let us focus on a neural network with a single associative pattern ($M=1$), which is similar to the Mattis model \cite{mattis1976solvable} and equivalent to a homogeneous mean-field Ising model \cite{kochmanski2013curie} (with energy $E(\bm x)=- \frac{1}{N} J\sum_{i<j} x_i x_j$) by changing a variable $x_i \leftarrow \xi_i x_i$. Rewriting \eqref{eq:mf-Hopfield}, we find that a one-pattern curved neural network follows a mean-field model given by
\begin{align}
    m &= \tanh \pr{\beta' J m},
    \label{eq:mf_one-pattern}
    \\ \beta' &= \frac{ \beta}{1+\gamma' \frac{1}{2} J m^2}.
    \label{eq:beta_one-pattern}
\end{align}
This result generalises the well-known Ising mean-field solution $m=\tanh\pr{\beta Jm}$, which is recovered for $\gamma=0$. 

By evaluating these equations, one finds that the model exhibits the usual order-disorder phase transition for positive and small negative values of $\gamma'$ (Fig.~\ref{fig:explosive-phase-transitions}.a top). However, for large negative values of  $\gamma'$, a different behaviour emerges: an explosive phase transition \cite{d2019explosive} that displays hysteresis due to HOIs (Fig.~\ref{fig:explosive-phase-transitions}.a bottom). The resulting phase diagram (Fig.~\ref{fig:explosive-phase-transitions}.b) closely resembles phase transitions in higher-order contagion models \cite{iacopini2019simplicial,landry2020effect} and higher-order synchronisation observed in Kuramoto models \cite{skardal2020higher}.

One can intuitively interpret the effect of the deformation parameter $\gamma'$ by noticing that, for a fixed $\beta'$, $m$  is the solution of a function of $\beta'$. For $\gamma'=0$, this results in the mean-field behaviour of the regular exponential model, which assigns a value of $m$ to each inverse temperature $\beta=\beta'$. 
In the case of the deformed model, the possible pairs of solutions $(m,\beta')$ are the same, but their mapping to the inverse temperatures $\beta$ changes. Namely, this deformation can be interpreted as a stretching (or contraction) of the effective temperature, which maps each pair $(m,\beta')$ to an inverse temperature $\beta=\beta'(1+\frac{1}{2}\gamma'Jm^2)$ according to \eqref{eq:beta_one-pattern}. Thus, one can obtain the mean-field solutions of the deformed patterns as mappings of the solutions of the original model. This is illustrated in Fig.~\ref{fig:explosive-phase-transitions}.c, where the solution of $\beta',m,\beta$ is projected to the planes $\beta=0$ and $\beta'=0$, obtaining the solutions for the flat ($\gamma'=0$) and the deformed ($\gamma'=-1.2$) models respectively.

In order to gain a deeper understanding of the explosive nature of this phase transition, we study the dynamics of the single-pattern neural network. By rewriting \eqref{eq:dynamical-mean-field} for $M=1$, and under the change of variables mentioned above to remove $\bm \xi$, the dynamical mean-field equation of the system reduces to
\begin{equation}
    \dot m=-m + \tanh\pr{\beta'J m},
\end{equation}
where $\beta'$ is calculated as in \eqref{eq:beta_one-pattern}. Simulations of the dynamical mean-field equations for values of $\beta$  just above the critical point are depicted in Fig.~\ref{fig:explosive-phase-transitions}.d. Trajectories with strongly negative $\gamma'$ saturate earlier than smaller negative $\gamma'$, confirming accelerated convergence. During this process, the effective inverse temperature $\beta'$ rapidly increases until it saturates, creating a positive feedback loop between $\beta'$ and $m$ that gives rise to the explosive nature of the phase transition. 
This positive loop occurs only if  $\gamma'$ is negative; otherwise, negative feedback simply makes the convergence of $m$ slower.

\subsection*{Overlaps between memory basins of attraction} 

\begin{figure*}
    \includegraphics[width=17.5cm]{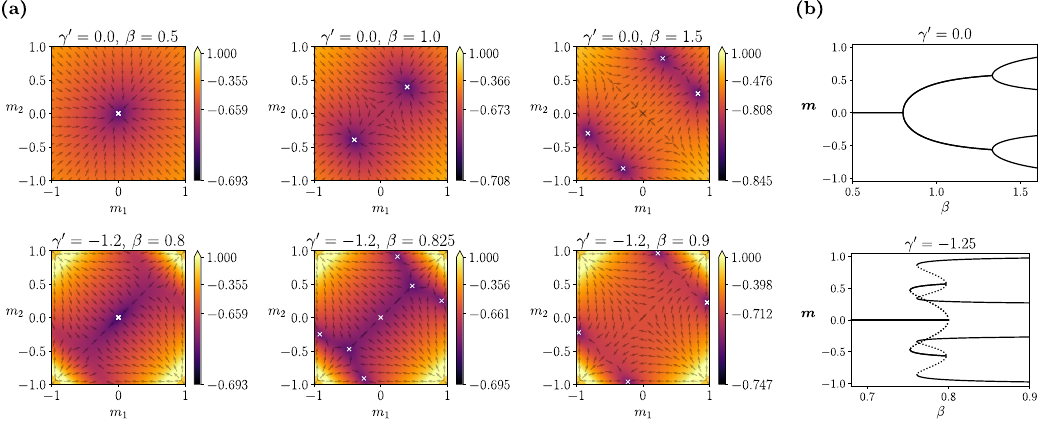}
    \caption{\textbf{Interaction between two encoded memories.}  \textbf{(a)} Values of $\varphi_{\gamma}$ for different mean-field values $m_1,m_2$, indicating the attractor structure of the network for different values of $\beta$ with $J=1, C=0.2$ for $\gamma'=0$ (top row) and $\gamma'=-1.2$ (bottom row). \textbf{(b)} Bifurcations of the order parameters $m_1,m_2$. For $\gamma'=0$ we observe an attractor bifurcating into two and then into four. For $\gamma'=-1.2$, we observe the same sequence, but with a coexistence hysteresis regime in which 7 attractors are possible.}
    \label{fig:bivariate-Hopfield}
\end{figure*}

A key property of associative-memory networks is their ability to retrieve patterns in different contexts. In the case of one-pattern associative-memory networks, the energy function $E(\bm x) = - \frac{J}{N} \sum_{i<j} x_i \xi_i \xi_j x_j$ is a quadratic function with two minima at $\bm x = \pm \bm \xi$, which configure global attractors. Instead, a two-pattern associative-memory network has an energy function with four minima (if sufficiently separated), but their attraction basins can overlap when the patterns are correlated.

To study the degree of the overlap between pairs of patterns, we analyse solutions of \eqref{eq:mf-Hopfield} for a network with two patterns with correlation $\ang{\xi_i^1 \xi_i^2}=C$ (see Supplementary Note~3.3
for details). 
In this scenario, the system is  described by two mean-field patterns:
\begin{align}
    m_{a}  &= \frac{1}{2}(1+C)\tanh \pr{ \beta'J (m_1 + m_2)} 
    \nonumber\\& \,\quad + w \frac{1}{2}(1-C) \tanh \pr{ \beta' J (m_1 -  m_2)}
\end{align}
with $w= 3-2a = \pm 1$ for $a=1,2$ and
\begin{equation}
     \beta' = \frac{ \beta}{1+\gamma \frac{1}{2} J(m_{1}^2+m_{2}^2)}.
\end{equation}

Fig.~\ref{fig:bivariate-Hopfield} shows how the hysteresis effect and explosive phase transitions persist in the case of two patterns for $C=0.2$ with negative $\gamma'$. This example shows two consecutive, overlapping explosive bifurcations (going from 1 to 2, and then to 4 fixed points), creating a hysteresis involving 7 fixed points within a more compressed parameter range of $\beta$ than the classical case. Consequently, the memory-retrieval region for the four embedded memories expands. These results illustrate complex hysteresis cycles as well as an increased memory capacity for finite temperatures by negative values of $\gamma'$. This enhanced capability for memory retrieval is further investigated through the replica analyses in the next section.

\subsection*{Memory retrieval with an extensive number of patterns}
\label{sec:infinite-patterns}

\begin{figure}
    \includegraphics[width=8.5cm]{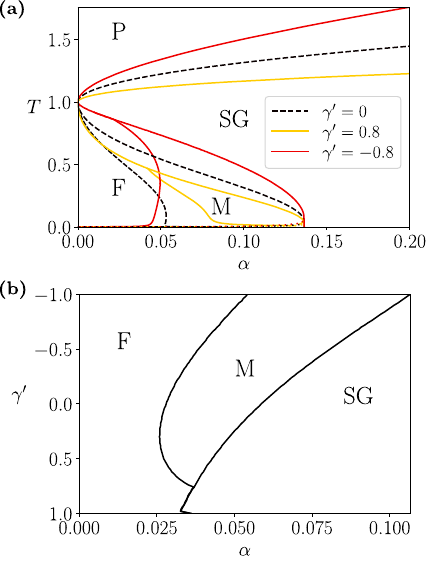}
    \caption{\textbf{Memory capacity is enhanced by geometric deformation.}
    Phase diagram of a curved associative memory with an extensive number of encoded patterns $M=\alpha N$ and $J=1$ for \textbf{(a)} different $T=1/\beta$ at $\gamma'=0$ (black dashed lines), $0.8, -0.8$ (solid lines), and for \textbf{(b)} different $\gamma'$ at $\beta=2$. F indicates the ferromagnetic (i.e., memory retrieval) phase, SG the spin-glass phase (where saturation makes memory retrieval inviable), M a mixed phase, and P the paramagnetic region. Both in F and M, ferromagnetic and spin-glass solutions coexist, but we differentiate these by calculating respectively whether memory-retrieval or spin-glass solutions are the global minimum of the normalising potential $\varphi_\gamma$. The dotted lines in (a) near $T=0$ indicate the AT lines, below which the replica-symmetric solution is not valid.
    Increasing $\gamma'$ to larger negative values extends the retrieval phase into larger values of $\alpha$, indicating an increased memory capacity, while larger positive values reduce the extension of the mixed phase, increasing robustness of memory retrieval.}
    \label{fig:memory-capacity}
\end{figure}

Next, we investigate how the deformation related to $\gamma$ impacts the memory-storage capacity of associative memories. In classical associative networks of $N$ neurons, the energy function is defined as $E(\bm x) = -\frac{J}{N}  \sum_{a=1}^M \sum_{i<j} x_i \xi_i^a \xi_j^a x_j$ with $M=\alpha N$. As the number of patterns learned by the network increases, the system transitions to a disordered spin-glass state in the thermodynamic limit. Furthermore, one can analytically solve this model \cite{amit1985storing,bovier1995gibbs,talagrand1998rigorous,shcherbina1993free}. For example, using the replica-trick method can determine the memory capacity of the system~\cite{amit1985storing}, and theoretically identify the critical value of $\alpha$ at which memory retrieval becomes impossible --- leading to a disordered spin-glass phase. Here, we apply a similar approach to reveal how deformed associative memory networks afford an enhanced memory capacity.

Applying the replica trick in conjunction with the methods outlined in previous sections allows us to solve the system (see Supplementary Note 5). 
This method entails computing a mean-field variable $m$ corresponding to one of the patterns $\bm\xi^a$ and averaging over the others. 
For simplicity, a pattern with all positive unity values $\bm\xi^a =(1,1,\ldots, 1)$ is considered, which is equivalent to any other single pattern just by a series of sign flip variable changes. 
The degree of similarity or overlap of this pattern with other patterns in the system introduces a new order parameter $q$, which contributes to measuring disorder in the system. 
After introducing the relevant order parameters and solving under a replica-symmetry assumption, the normalising potential is derived as 
\begin{align}
    \varphi_{\gamma} &= 
     N\frac{\beta}{\gamma'} \ln{\frac{\beta}{\beta'}}
      -  N\beta' J m^2 - N\frac{1}{2} \alpha (\beta' J)^2 (r +R-2qr)  
    \nonumber \\& \,\quad -N\frac{1}{2} \alpha \pr{ \ln\pr{1-\beta'J(1-q)} -\beta'J\sqrt{rq}}
    \nonumber \\& \,\quad + N \int Dz \ln \pr{ 2 \cosh\pr{ \beta' J m +  \beta' J \sqrt{\alpha r} z }}, 
    \label{eq:phi_near_sat}
\end{align}
where $J$ is a scaling factor, and the order parameters are defined as
\begin{align}
     m &= \int Dz\tanh \pr{ \beta'  J m +  \beta' J \sqrt{\alpha r} z } \label{eq:m_near_sat},
 \\q &= \int Dz \tanh^2\pr{\beta' J m  + \beta' J \sqrt{\alpha r} z  } \label{eq:q_near_sat}
\end{align}
with
\begin{align}
    r= \frac{q}{(1 -\beta'J (1 - q))^2} , \quad R = \frac{(\beta' J)^{-1} - (1-2 q)}{(1 - \beta'J (1 - q))^2} .
\end{align}
As in previous cases, the model is governed by an effective temperature
\begin{equation}
    \beta' = \frac{\beta}{1+\gamma' \frac{1}{2} \pr{ J m^2 + \alpha  J( \beta' (R - qr) - 1)}}.\label{eq:beta_RS}
\end{equation}
This solution differs from the models in previous sections by the self-dependence of $\beta'$.

To obtain a phase diagram, we solved (\ref{eq:m_near_sat}-\ref{eq:q_near_sat}) numerically for given $\alpha,\beta'$ at $\gamma'=0$, and rescaled the inverse temperature as in the previous section to obtain the corresponding values of $\beta$ for each $\gamma'$. Using the resulting order parameters and calculating the free energy for each $\alpha,\beta,\gamma'$, we constructed the phase diagram of the system (similarly to \cite{amit1985storing,coolen2001statistical}) characterised by the following distinct phases (Fig.~\ref{fig:memory-capacity}):
\begin{itemize}
    \item A paramagnetic phase (P), corresponding to disordered solutions with $m=q=0$, where memory-retrieval fails due to the dominance of fluctuations. 
    \item A ferromagnetic phase (F), corresponding to stable memory-retrieval solutions with $m > 0$ and $q > 0$.
    \item A spin-glass phase (SG), exhibiting spurious-retrieval solutions with $m = 0$ and $q > 0$. 
    \item A  mixed phase (M), where F and SG types of solutions coexist, being the spin-glass solutions a global minimum of the normalising potential $\varphi_\gamma$. 
\end{itemize}
For $\gamma' = 0$ (black dashed lines), the phase transition reflects the behaviour of associative memories near saturation \cite{amit1985storing,coolen2001statistical}. With negative $\gamma'$ (red lines), we observe an expansion of the ferromagnetic and mixed phases, indicating an enhanced memory-storage capacity by the deformation. Conversely, a positive value of $\gamma'$ (yellow lines) decreases the memory capacity but reduces the extent of the mixed phase. In the mixed phase, retrieved memories ($m>0$) are represented at a local --- but not global --- minimum of the normalising potential $\varphi_{\gamma}$ in \eqref{eq:phi_near_sat}, indicating a larger probability of observing spurious patterns. Thus, we expect positive values of $\gamma'$ to result in more robust memory retrieval.

The stability of the replica symmetry solution is given by the condition
\begin{equation}
    \big(1+ \beta' (1-q)\big)^2 > \alpha \beta'^{2}  \int Dz \cosh^{-4} \beta' \pr{ J m + J \sqrt{\alpha r} z },
    \label{eq:RS_cond}
\end{equation}
which is captured by the dotted lines near zero temperature in Fig.~\ref{fig:memory-capacity}.a. Note that all solutions in Fig.~\ref{fig:memory-capacity}.b are stable under the replica symmetry assumption.

\begin{figure}
    \includegraphics[width=8.5cm]{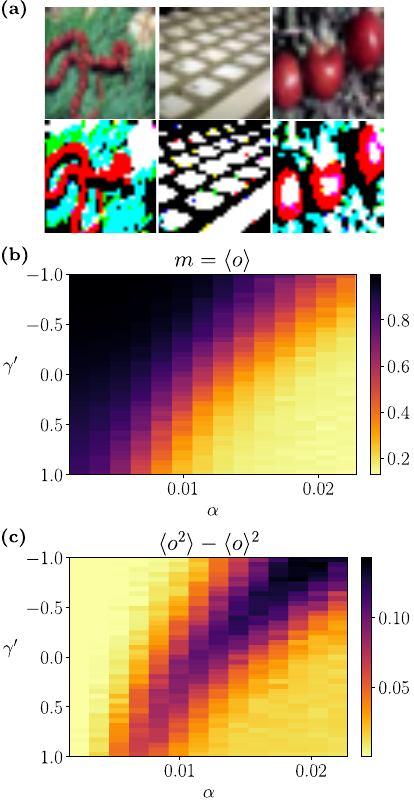}
    \caption{\textbf{Simulation study for the effect of deformation on image encoding.}
    \textbf{(a)} Examples of CIFAR-100 images (top) and their RGB binarised versions (bottom). Every 32x32x3 binary RGB pixel value for each image $a$ is assigned to the value of one position of pattern $\xi_i^a$. \textbf{(b, c)} Mean and variance of pattern retrieval values obtained in experiments, measured by the overlap between the final state of the network and the encoded pattern.}
    \label{fig:cifar}
\end{figure}

We complement the analysis from the previous section with an experimental study of a system encoding patterns from an image classification benchmark. The patterns are sourced from the CIFAR-100 dataset, which comprises $60,000$ 32x32 colour images~\cite{krizhevsky2009learning}.
To adapt the dataset to binary patterns suitable for storage in an associative memory, we processed each RGB channel by assigning a value of $1$ to pixels with values greater than the channel's median value and $-1$ otherwise (Fig.~\ref{fig:cifar}.a). The resulting array of $N=32\cdot 32\cdot 3$ binary values for each image was assigned to patterns $\bm\xi^a$. Note that associative memories (as well as our theory above) usually assume that patterns are relatively uncorrelated, and specific methods are required to adapt them to correlated patterns \cite{fontanari1990storage,agliari2013parallel}. To simplify the problem, we conducted experiments using a selection of 100 images with covariance values smaller than $10/\sqrt{N}$ (the standard deviation of the covariance values for uncorrelated patterns is $1/\sqrt{N}$). We used a random search to select patterns with low correlations: we randomly picked an image and replaced it if its correlation exceeded the threshold, repeating until all correlations were below it.

We evaluated the memory retrieval capacity of networks with various degrees of curvature $\gamma$ by encoding different numbers of memories, 
as described in \eqref{eq:weight-encoding}. As a measure of performance, we evaluated the stability of the network by assigning an initial state $\bm x = \bm \xi^a$ and calculating the overlap $o = \sum_i x_i \xi_i^a$ after $T=30N$ Glauber updates for $\beta=2, J=1$. The process was repeated $R=500$ times from different initial conditions (different encoded patterns and different initial states) to estimate the value of $m$ in \eqref{eq:m_near_sat}. Experimental outcomes confirm our theoretical results, revealing that memory capacity increases with negative values of $\gamma'$, while positive values reduce the memory capacity (Fig.~\ref{fig:cifar}.b), but reduce the extent and magnitude of the high variability region in pattern retrieval (Fig.~\ref{fig:cifar}.c), which is consistent with the reduction of the mixed phase. Note that the resulting memory capacity of the system observed in our experiments (i.e., the value of $\alpha$ at which the transition happens) is diminished due to the presence of correlations among some of the memorised patterns.

\begin{figure}
    \centering
    \includegraphics[width=8.5cm]{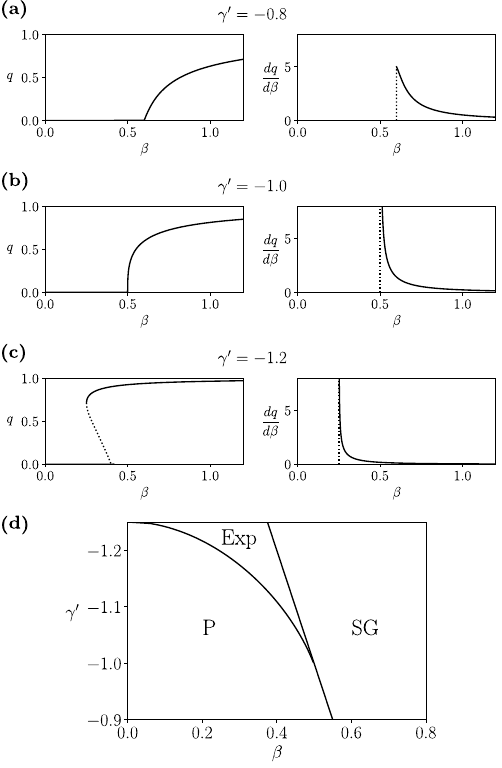}
    \caption{\textbf{Explosive spin glasses.}
    Phase transitions for order parameter $q$ for replica-symmetric disordered spin models displaying \textbf{(a)} a cusp phase transition for $\gamma'=-0.5$, \textbf{(b)} a second-order phase transition for $\gamma'=-1.0$ and \textbf{(c)} an explosive phase transition for $\gamma'=-1.2$. \textbf{(d)}~Phase diagram of the explosive spin glass, displaying a paramagnetic (P), spin-glass (SG) and an explosive phase (Exp).}
    \label{fig:spin-glass}
\end{figure}

Finally, we investigated transitions near the spin-glass phase boundaries. First, we note that, for $J\to 0$ and $\alpha=J^{-2}$, the model in (\ref{eq:m_near_sat}-\ref{eq:q_near_sat}) converges to (see Supplementary Note~5) 
\begin{align}
    q &= \int Dz \tanh^2\pr{\beta' \sqrt{q} z},
    \\   \beta' &= \frac{\beta}{1+\frac{1}{2}\gamma' \beta' (1- q^2) } ,
    \label{eq:replica-symmetry-solution-SK}
\end{align}
which at $\gamma=0$ recovers the well-known Sherrington-Kirkpatric model 
\cite{sherrington1975solvable} (see Supplementary Note~6).
While in the classical case, a phase transition occurs from a paramagnetic to a spin-glass phase, the curvature effect of $\gamma'\neq 0$ modifies the nature of this transition. For small values of $\gamma'$, the system exhibits a continuous phase transition akin to the Sherrington-Kirkpatrick spin-glass, where $\frac{dq}{d\beta}$ shows a cusp (Fig.~\ref{fig:spin-glass}.a). However, for $\gamma'=-1$ the phase transition becomes second-order, displaying a divergence of $\frac{dq}{d\beta}$ at the critical point (Fig.~\ref{fig:spin-glass}.b). Moreover, increasing the magnitude of negative $\gamma'$ leads to a first-order phase transition with hysteresis (Fig.~\ref{fig:spin-glass}.c), resembling the explosive phase transition observed in the single-pattern associative-memory network. This hybrid phase transition combines the typical critical divergence of a second-order phase transition with a genuine discontinuity, similar to `type V' explosive phase transitions as described in \cite{d2019explosive}.

We analytically calculated the properties of these phase transitions (see Supplementary Note~6).  
By computing the solution at $\gamma'=0$ and rescaling $\beta'$, we determined that the critical point is located at $\beta_c = 1 + \frac{1}{2}\gamma'$ (consistent with Fig.~\ref{fig:spin-glass}.a-c). The slope of the order parameter around the critical point is, for $\gamma' \leq -1$, equal to $(1+\gamma')^{-1}$, indicating the onset of a second-order phase transition as depicted in Fig.~\ref{fig:spin-glass}.b. The resulting phase diagram of the curved Sherrington-Kirkpatrick model is shown in Fig.~\ref{fig:spin-glass}.d.

\subsection*{Comparison with other dense associative memory models}

Although our primary objective is to develop a parsimonious model of HOIs to explain higher-order phenomena, our framework can also be used to explain the behaviour of modern networks with HOIs, including the recently proposed relativistic Hopfield model~\cite{barra2018new,agliari2019relativistic,agliari2020generalized} and dense associative memories \cite{krotov2016dense,demircigil2017model}. For this, let us consider the energy $\mathcal{F}[E]$ of the exponential family distribution $p(\bm x) \sim e^{-\beta \mathcal{F}[E]}$ given by the nonlinear transformation (denoted by $\mathcal{F}$) of the classical energy $E(\bm x)$. 
The deformed exponential models in this study correspond to $\mathcal{F}[E]  = - \frac{ N}{\gamma'}  \ln(1-\gamma' E/N)$, while the relativistic model corresponds to \(\mathcal{F}[E] =- \frac{ N}{\gamma'} \sqrt{1-\gamma' E/N}\). 
For the deformed exponential, the term $\mathcal{F}[E]$ can be expanded as 
\begin{align}
    \mathcal{F}[E]
    &= E +\frac{\gamma'}{2N} E^2 
    +\frac{\gamma'^2}{3N^2} E^3 + \dots 
\end{align}
When $E$ depends on the quadratic Mattis magnetisation (i.e., $E=- \sum_a \frac{1}{N}\pr{\sum_i \xi_i^a x_i}^2$), then $\mathcal{F}[E]$ expands in terms of even-order HOIs of $\sum_i \xi_i^a x_i$. For $\gamma'<0$, all coefficients of $\sum_i \xi_i^a x_i$ in the expansion are negative, indicating that embedded memories have deeper energy minima than in the classical case. The same signs appear for each order in the relativistic energy with $\gamma'<0$. 
We also note that $\beta$ in the free energy of both the deformed exponential and relativistic models in the limit of large $N$ appears scaled according to an effective temperature given by $\beta' = \beta \partial_{E} \mathcal{F}[E]$  (e.g., \eqref{eq:neg_norm_potential} and Eq.~(6.2) in Ref.~\cite{agliari2020generalized}). Moreover, the input in the Glauber dynamics is approximated for large sizes as 
\begin{align}
    \beta \Delta \mathcal{F}[E] \approx \beta \partial_{E} \mathcal{F}[E] \, \Delta E(\bm x)=\beta' \Delta E(\bm x).
\end{align}
The effective inverse temperatures $\beta' =  \beta (1 - \gamma'  E/N)^{-1}$ for the deformed exponential and $\beta' = 2^{-1}(1 - \gamma'  E/N)^{-1/2}$ for the relativistic models are decreasing functions of $E$ when $\gamma'<0$, resulting in an acceleration of memory retrieval --- with lower energy $E$ resulting in higher $\beta'$ (lower temperature).
While the relativistic model has been studied for $\gamma'>0$~\cite{barra2018new,agliari2019relativistic,agliari2020generalized}, we conjecture it may exhibit explosive phase transitions if $\gamma'<0$.
Conversely, a positive $\gamma'$ introduces alternating signs in even-order terms of $\sum_i \xi_i^a x_i$, and a shallower energy landscape due to a reduction in $\beta'$. This shallower energy landscape reduces the memory capacity of the deformed exponential networks by expanding the spin-glass phases (Fig.~\ref{fig:memory-capacity}), but also enlarges the recall (ferromagnetic) region by mitigating the formation of spurious memories given by overlapping patterns in the mixed phase (See also \cite{barra2018new} for mitigation of spurious memories in the relativistic model). 

This perspective on accelerated memory retrieval by nonlinearity extends to dense associative memories~\cite{krotov2016dense,demircigil2017model}, which achieve supralinear memory capacities through nonlinear pattern encoding. Specifically, their energy function is given by $\mathcal{F} = -\sum_{a} F(\sum_i \xi_i^a x_i)$ with $F$ being e.g., a thresholded power function, $F(z)=\br{z}_+^p$~\cite{krotov2016dense} or an exponential nonlinearity $F(z)=e^{z}$~\cite{demircigil2017model} at zero temperature. These nonlinearities narrow basins of attraction, reducing memory overlap and preventing transitions to the spin-glass phase. 
The jumps in the Glauber dynamics of such systems are weighed by an accelerating function. Namely, from our perspective, the dynamics of such systems can be described via positive feedback on weights linked to a specific memory, which increase during memory retrieval. This follows from the fact that, relating the linear difference in Mattis terms $\Delta \epsilon_k^a \equiv 2 \xi_k^a x_k$ with the nonlinear difference $\Delta F_k^a \equiv F\left(\sum_i \xi_i^a x_i\right) -  F\left(\sum_{i} \xi_i^a x_i - \Delta \epsilon_k^a\right)$, the update of the $k$th neuron is determined by the sign of
\begin{align}
	\Delta \mathcal{F}(\bm{x})
    = \sum_a  \frac{\Delta F_k^a}{\Delta \epsilon_k^a} \Delta \epsilon_k^a = \sum_a w_{k}^{a} \Delta \epsilon_k^a. 
\end{align}
Here, we show that the effective weight 
$w_{k}^{a} \equiv \frac{\Delta F_k^a}{\Delta \epsilon_k^a}$ becomes an increasing function of $\sum_i \xi_i^a x_i$ when $F$ is the power, exponential, or more generally, a convex function (See Supplementary Note~7). 
Thus, increasing $\sum_i \xi_i^a x_i$ as pattern $\bm \xi^a$ is retrieved strengthens its basin of attraction and ensures positive feedback. Meanwhile, retrieval of $\bm \xi^a$ reduces $\sum_i \xi_i^{b} x_i$ for orthogonal patterns $\bm \xi^b$, lowering their weights, suppressing their recall to minimize interference. This competitive mechanism highlights the higher memory capacity of these models compared to curved neural networks with uniform temperature scaling. Unlike the effective inverse temperature in curved networks, which depends only on the system's state or energy, the effective weight in updating $k$-th neuron additionally depends on the neuron's state $x_k$, thus no longer representing a global modulation of the energy.

\section*{Discussion}

HOIs play a critical role in enabling emergent collective phenomena in natural and artificial systems. Modelling HOIs is, however, highly non-trivial, often requiring advanced analytic tools (such as simplicial complexes or hypergraphs) that entail an exponential increase in parameters for large systems. In this paper, we addressed this issue by leveraging the maximum entropy principle to effectively capture HOIs in models via a deformation parameter $\gamma$, which is associated with the Rényi entropy. Given their close connection with statistical physics, this family of models provides a useful setup to investigate the effect of HOIs on spin systems, including explosive ferromagnetic and spin-glass phase transitions, extending studies on anomalous phase transitions found in other systems ~\cite{angst2012explosive,iacopini2019simplicial,landry2020effect,battiston2021physics,d2019explosive}, and the capability of networks to store memories.

The observed effects in curved neural networks can be explained via an effective temperature, inducing a positive or negative feedback effect in memory retrieval. As we discussed above, this effect is present in different forms across other dense associative memories \cite{agliari2020generalized,krotov2016dense,demircigil2017model}. A similar argument may apply to diffusion models framed within dense associative memories~\cite{ambrogioni2024search,ambrogioni2023statistical}, where the energy follows a log-sum-exp nonlinearity. Thus, the accelerated mechanism found in this study clarifies memory retrieval in advanced associative networks, providing an important step toward designing extended memory capacities and improved noise scheduling.

Curved neural networks also provide insights into biological neural systems, where evidence suggests the presence of alternating positive and negative HOIs for even and odd orders, respectively. This alternation leads to sparse neuronal activity, which has been shown to be instrumental for enabling extended periods of total silence~\cite{ohiorhenuan2010sparse,tkavcik2014searching,shimazaki2015simultaneous,shomali2023uncovering,rodriguez2023alternating}. Interestingly, such sparse activity patterns may coexist with the accelerated memory retrieval dynamics, as both involve positive even-order HOIs. The attainment of enhanced memory, combined with sparse activity, presents a promising direction for understanding energy-efficient biological neuronal networks~\cite{rodriguez2023alternating,santos2024hopfield}. 
Future work may investigate how curved neural networks might support both energy efficiency and high memory capacities, potentially by adopting a thresholded, supralinear neuronal activation function \cite{krotov2016dense, rodriguez2023alternating}. Additionally, developing statistical methods for fitting these models to experimental data (i.e., theories for learning) represents an important, yet largely unexplored, research avenue. Together, these research directions offer a compelling path to uncover the principles of efficient information coding in biological neural systems.

Overall, our results demonstrate the benefits of considering the maximum entropy principle, emergent HOIs, and nonlinear network dynamics as theoretically intertwined notions. As showcased here, such an integrated framework reveals how information encoding, retrieval dynamics, and memory capacity in neural networks are mediated by HOIs, providing principled, analytically tractable tools and insights from statistical mechanics and nonlinear dynamics. 
More generally, the framework presented in this work extends beyond neural networks and contributes to a general theory of HOIs, paving the road toward a principled study of higher-order phenomena in complex networks.

\section*{Data Availability}
The CIFAR-100 dataset used in this study is available at \url{https://www.cs.toronto.edu/~kriz/cifar.html}

\section*{Code Availability}
The code generated in this study is available in the GitHub repository, \url{https://github.com/MiguelAguilera/explosive-neural-networks}.

\def\bibsection{\section*{References}}
\bibliography{references}

\begin{acknowledgments}
We thank Ulises Rodriguez Dominguez for valuable discussions on this manuscript. 
M.A. is funded by a Junior Leader fellowship from `la Caixa' Foundation (ID 100010434, code LCF/BQ/PI23/11970024), John Templeton Foundation (grant 62828), Basque Government ELKARTEK funding (code KK-2023/00085) and Grant PID2023-146869NA-I00 funded by MICIU/AEI/10.13039/501100011033 and cofunded by the European Union, and supported by the Basque Government through the BERC 2022-2025 program and by the Spanish State Research Agency through BCAM Severo Ochoa excellence accreditation CEX2021-01142-S funded by MICIU/AEI/10.13039/501100011033.
P.A.M. acknowledges support by JSPS KAKENHI Grant Number 23K16855, 24K21518. 
F.R. is supported by the UK ARIA Safeguarded AI programme and the PIBBSS Affiliatership programme. 
H.S. is supported by JSPS KAKENHI Grant Number JP 20K11709, 21H05246, 24K21518, 25K03085. 
\end{acknowledgments}

\section*{Author Contribution}
M.A., P.A.M, F.E.R and H.S. designed and reviewed the research and wrote the paper. M.A. contributed the analytical and numerical results. P.A.M. contributed part of the analytical results of the replica analysis.

\section*{Competing interests}
The authors declare no competing interests.

\clearpage

\appendix
\onecolumngrid

\begin{center}
    
{\large Explosive neural networks via higher-order interactions in curved statistical manifolds}
\vspace{1em} \\
{\large \textbf{Supplementary Information}}
\vspace{1em} \\
Miguel Aguilera \\
\textit{BCAM – Basque Center for Applied Mathematics, 
        Bilbao, Spain \\ 
        IKERBASQUE, Basque Foundation for Science, Bilbao, Spain}
\vspace{1em} \\
Pablo A. Morales \\
    \textit{Research Division, Araya Inc., Tokyo, Japan \\
    Centre for Complexity Science, Imperial College London, London, UK}
\vspace{1em} \\
Fernando E. Rosas\\
    \textit{Sussex AI and Sussex Centre for Consciousness Science,\\
    Department of Informatics, University of Sussex, Brighton, UK\\
    Department of Brain Science and Centre for Complexity Science, Imperial College London, London, UK\\
    Center for Eudaimonia and Human Flourishing, University of Oxford, Oxford, UK\\
    Principles of Intelligent Behavior in Biological and Social Systems (PIBBSS), Prague, Czech Republic}
\vspace{1em} \\
Hideaki Shimazaki \\
    \textit{Graduate School of Informatics, Kyoto University, Kyoto, Japan\\
    Center for Human Nature, Artificial Intelligence,\\
    and Neuroscience (CHAIN),  Hokkaido University, Sapporo, Japan}
\end{center}

\renewcommand\appendixname{Supplementary Note}
\setcounter{figure}{0}
\renewcommand{\thefigure}{S\arabic{figure}}
\renewcommand{\thesection}{\arabic{section}}
\renewcommand{\theequation}{S\thesection.\arabic{equation}}
\setcounter{page}{1}

\section{Maximum Rényi entropy and information geometry}
\label{app:max-renyi-ent}

The maximum entropy principle (MEP) is a framework for building parsimonious models consistent with observations, being particularly well-suited for the statistical description of systems in contexts of incomplete knowledge~\cite{jaynes2003probability,cofre2019comparison}. The MEP uses entropy as a fundamental tool to quantify the degree of structure present in a given model. 
Accordingly, the MEP suggest to adopt the model with the maximal entropy --- i.e. the least amount of structure --- that is consistent with selected features of the data (for example, their first- and second-order statistics), following the idea that no additional regularities should be introduced beyond the ones specified by those.

Maximum entropy models are particularly well-suited for the study of neural systems. 
By abstracting neurons into binary variables $x_i$ representing the presence or absence of action potentials, the MEP provides a powerful approach to model collective neural activity. 
In this approach, the Ising model emerges from the maximisation of Shannon entropy under constraints on activity rates of individual neurons and pairwise correlations:
\begin{equation}
    p^{(2)} = \argmax_{q(\bm x)} H(\bm x) 
    \quad \text{s.t.}\quad 
    \begin{cases}
    \ang{x_i} &= \eta_i, \\
    \ang{x_i x_j} &= \eta_{ij}, 
    \end{cases}
    \nonumber
\end{equation}
where $H(\bm x)$ denotes the Shannon entropy of $\bm x=\{x_1,\dots,x_n\}$ under distribution $q(\bm x)$. 
It can be shown that
\begin{equation}
    p^{(2)}(\bm x) = \frac{1}{Z}\exp\pr{\sum_i \theta_i x_i + \sum_{i<j} \theta_{ij} x_ix_j},
\end{equation}
with $Z$ being a normalising constant. Hence, this model encapsulates observed information up to second-order statistics, represented in how $\theta_i,\theta_{ij}$ depend on the constraints $\eta_i,\eta_{ij}$. 
Furthermore, the dynamics of the Ising model can be investigated via exact solutions, approximations (encompassing mean-field and Bethe approximations), and simulations, thereby providing a rich set of insights and analytical tools. 
The Ising model has been instrumental in the development of recurrent neural networks, leading to Hopfield networks and Boltzmann machines.

What if the observations that one is to model require us to consider statistics beyond pairwise interactions?
Following the same principle, one can construct models with third- and higher-order interactions~\cite{amari2001information} resulting in distributions of the following type:
\begin{equation}
\label{eq:gen_ising}
    p^{(k)}(\bm x) = \frac{1}{Z}\exp\vast( \sum_{\substack{\bm I\subseteq\{1,\dots,n\}\\|\bm I|\leq k}} \theta_{\bm I} \prod_{i\in\bm I} 
    x_i \vast)
\end{equation}
with the summation going over all subsets of $k$ or less variables. 
Above, the argument within the exponential is an energy function $E_k(\bm x)$, with the index $k$ highlighting the highest order of interactions considered. 
It is important to notice that the number of terms in the Hamiltonian grows exponentially with $k$, making it unfeasible in practice to construct models including high orders $k\gg 1$.

\subsection{Capturing high-order interactions via non-Shannon entropies}

While traditional formulations of the MEP are based on Shannon's entropy~\cite{jaynes1957information}, more recent work has expanded it to include other entropy functionals, including the entropies of Tsallis~\cite{tsallis1998role} and Rényi~\cite{morales2021generalization}. 
Here we argue that some high-order interdependencies can be efficiently captured by the deformed exponential family~\eqref{eq:deformed_exp}, which arises as a solution to the problem of maximising non-Shannon entropies --- as explained below. 

By starting from a conventional MEP model with few degrees of freedom tuned to account for low-order interactions, one can enhance its capability to account for higher-order interdependencies by the inclusion of a deformation parameter, defined as an extension of the Rényi's index (or Tsallis's $q$ or Amari's $\alpha$), with clear geometrical interpretation, i.e. the scalar curvature of the manifold~\cite{morales2021generalization}. Concretely, let's consider the Rényi entropy with parameter $\gamma\geq -1$, given by
\begin{equation}
    H_\gamma = - \frac{1}{\gamma} \ln \sum_{\bm x} p(\bm x)^{1+\gamma}.
\end{equation}
This definition adopts the shifted indexing convention introduced in Ref.~\cite{valverde2019case}, thereby referring to $\gamma = \alpha-1$ as the order of Rényi's entropy, with $\alpha\ge 0$ corresponding to the order in the standard definition.
Rényi entropy recovers the standard Shannon entropy at the limit $\gamma \to 0$. 
The maximisation of the Rényi entropy can be performed by extremisation of the Lagrangian:
\begin{equation}
    \mathcal{L} 
    = - \frac{1}{\gamma} \ln \sum_{\bm x} p(\bm x)^{1+\gamma} + \theta_0 \left(\sum_{\bm x} p(\bm x) - 1\right)  
     +\theta_0 \gamma\beta\sum_{i=1}^L \theta_i\left(\sum_{\bm x} p(\bm x) f_i(\bm x)-c_i\right), 
\end{equation}
which also consider constraints $\sum_{\bm x} p(\bm x) =1$ and $\sum_{\bm x} p(\bm x) f_i(\bm x)=c_i$ with $i=1,\dots,L$, whereas the first ensures $p(\bm x)$ to be a probability mass function and the second fixes the average of $f_i(\bm x)$ on a desired value $c_i$. 
Note that the coefficient $\beta$ is introduced to keep $\gamma$ dimensionless, corresponding to the inverse temperature in statistical physics. This results in the maximum entropy condition
\begin{equation}
    0=\frac{\delta\mathcal{L}}{\delta p(\bm x)} = - \frac{1}{\gamma} \frac{ (1+\gamma)p(\bm x)^{\gamma} }{\sum_{\bm x} p(\bm x)^{1+\gamma}}  + \theta_0  + \theta_0 \gamma \beta\sum_a \theta_a f_a(\bm x).
\end{equation}
The family of probability distributions meeting the above condition is known as the \emph{deformed exponential family}, which is given by
\begin{equation}
\label{eq-app:deformed_exp}
    p_{\gamma}(\bm x) = \exp(-\varphi_{\gamma})  \bigg[1+ \gamma  \beta\sum_a \theta_a f_a(\bm x)\bigg]_+^{1/\gamma}
\end{equation}
where $\varphi_{\gamma}$ is a normalising constant
\begin{equation}
    \varphi_{\gamma} = \ln \sum_{\bm x} \bigg[1+ \gamma  \beta\sum_a \theta_a f_a(\bm x)\bigg]_+^{1/\gamma} .
\end{equation}
Above, we use the square bracket $\brpos{\cdot}$ operator to set negative values to zero, so that $\brpos{x}=\max\{0,x\}$. In the next sections, to solve the steepest descent step of mean field calculations, we will assume that the content of the  $\brpos{\cdot}$ operator is always possible. This assumption is reasonable under an adequate normalisation of $\gamma$.

Importantly, Rényi's entropy is closely related to Tsallis' entropy
\begin{equation}
    H_\gamma^{(\mathrm{Ts})} = - \frac{1}{\gamma} \pr{1- \sum_{\bm x} p(\bm x)^{1+\gamma}}.
\end{equation}
It can be shown that the Tsallis and Rényi's entropies can be deformed into one another by a monotonically increasing function. This fact brings both divergences, from the geometrical perspective, to the same equivalence class generating the same geometry, see Ref.~\cite{morales2021generalization}. 
In particular, by maximising Tsallis entropy, one recovers the same deformed exponential family, $p_{\gamma}$, using $q = 1 - \gamma$~\cite{umarov2008aq}.

We also note that maximising Rényi's entropy with constraints from the expectation given by the escort distribution leads to a similar distribution, but with the exponent replaced by $-1/\gamma$ (see Theorem 3.15 in \cite{wong2022tsallis}).

\newpage
\section{Glauber rule}
\label{app:Glauber}

Glauber dynamics is a Markov Chain Monte Carlo algorithm that is popular for simulating neural activity according to Hopfield networks and Ising models. In this method, one samples the activity of each neuron conditioned on the activity of other neurons according to the following conditional distribution:
\begin{equation}
	p_{\gamma}(x_k | \bm{x}_{\backslash k}) 
	= \frac{ p_{\gamma}(x_k, \bm{x}_{\backslash k}) }{ p_{\gamma}(\bm{x}_{\setminus k}) }  
	= \frac{ p_{\gamma}(x_k, \bm{x}_{\backslash k}) }{ p_{\gamma}(x_k, \bm{x}_{\backslash k}) + p_{\gamma}(-x_k, \bm{x}_{\backslash k}) }  
    = \frac{ 1 }{ 1 + \frac{ p_{\gamma}(-x_k, \bm{x}_{\backslash k}) } {p_{\gamma}(x_k, \bm{x}_{\backslash k})} }, 
\end{equation}
where $\bm{x}_{\backslash k}$ denotes the state of all neurons except the $k$-th one. This sampling procedure is carried out for all neurons in an iterative manner.

\begin{figure}[b]
    \centering
    \begin{minipage}{0.45\textwidth}
        \centering
        \includegraphics[width=\linewidth]{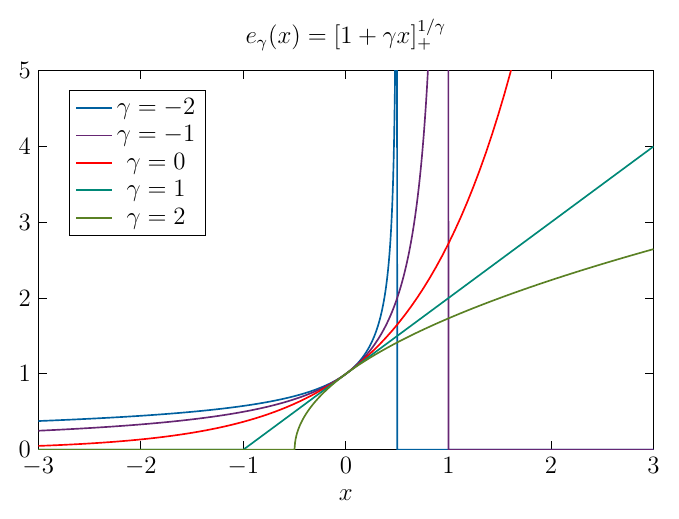} 
    \end{minipage}\hspace{0.02\textwidth} 
    \begin{minipage}{0.45\textwidth}
        \centering
        \includegraphics[width=\linewidth]{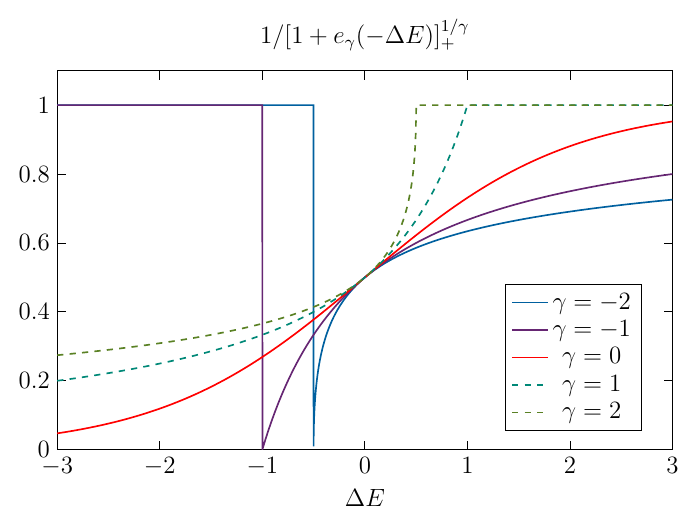} 
    \end{minipage}
    \caption{
    \textbf{(Left)} The deformed exponential functions, $e_{\gamma}(x) = [1 + \gamma x]_+^\frac{1}{\gamma}$. 
    \textbf{(Right)} The activation function of a neuron as a function of input $\Delta E$.}
    \label{fig:enter-label}
\end{figure}

Let us construct Glauber dynamics for a curved neural network. The deformed exponential family distribution states that the distribution of $\bm x$ is given by 
\begin{equation}
    p_{\gamma}(\bm x) = \exp\pr{-\varphi_{\gamma}}\brpos{1- \gamma \beta E(\bm x)
    }^{1/\gamma},
\end{equation}
where the energy function $E(\bm x)$ is given by
\begin{equation}
    E(\bm x) = - \sum_{i} H_i x_i -  \frac{1}{2N}\sum_{i,j} J_{ij} x_i x_j
\end{equation}
with $J_{ii}=0$ and $J_{ij}=J_{ji}$. The deformed exponential family distribution can be rewritten as
\begin{equation}
    p_\gamma(-x_k, \bm{x}_{\backslash k}) =\exp\pr{-\varphi}\brpos{1- \gamma \beta \pr{E(\bm x) + 2x_k h_k }}^{1/\gamma}
\end{equation}
with $h_k = H_k + \frac{1}{N}\sum_j J_{kj} x_j$. 
Under the assumption of $1- \gamma \beta E(\bm x) >0$ (and the same for the state resulting from flipping the $k$-th spin), a direct derivation shows that
\begin{align}
    p_\gamma(x_k | \bm{x}_{\backslash k}) 
    &= \pr{1 + \pr{ \frac{1- \gamma \beta \pr{E(\bm x) + 2x_k h_k }}{1- \gamma \beta E(\bm x)  }} ^{1/\gamma} }^{-1}
    \nonumber \\ &= \pr{1 + \pr{1  -\gamma 2 \beta' x_k h_k}^{1/\gamma}}^{-1}
    \nonumber \\ &=  \pr{1 + \exp_\gamma\pr{-2\beta' x_k h_k}}^{-1}
    \\ \beta' &= \frac{\beta}{1-\gamma \beta E(\bm x)}.
\end{align}
where $\exp_\gamma(\cdot )$ stands for the deformed exponential. Note that these equations recover the classical Glauber rule for Ising models at $\gamma = 0$. 
Fig.~S1 shows the deformed exponential function and the activation function $p_\gamma(x_k=1| \bm{x}_{\backslash k})$ as a function of input $\Delta E = 2\beta' h_k$, representing the deformed nonlinearity of a neuron. Note that to have a smooth activation function, the input must satisfy $1+\gamma \Delta E>0$, resulting in $\Delta E>-1/\gamma$ if $\gamma>0$ and $\Delta E<-1/\gamma$ if $\gamma<0$. For implementing the sampling strategy, the selection of neurons can be sequential, using random permutations, or using probabilistic methods (according to non-zero probabilities assigned to each neuron). 

In the case of large systems in which $E(\bm x)$ is extensive, then a normalisation of the curvature parameter in the form $\gamma' = \frac{\gamma}{N}$ is required. This makes the value of $\gamma x_k h_k$ tend to zero as $N\to\infty$. 
In this case, calculating the limit of $\exp_\gamma\pr{- 2\beta' x_k h_k}$ as $\gamma\to 0$, one finds that
\begin{equation}
    p_\gamma(x_k | \bm{x}_{\backslash k})  =  \pr{1 + \exp\pr{-2\beta' x_k h_k}}^{-1} = \frac{\exp\pr{\beta' x_k h_k}}{2\cosh\pr{\beta' h_k}},
    \label{eq:glauber-approximation}
\end{equation}
with effective temperature $\beta'$ given by    
\begin{equation}    
\beta' = \frac{\beta}{1-\gamma'\frac{1}{N} E(\bm x)}.
\end{equation}

\newpage
\section{The mean-field theory of curved neural network}
\label{app:hopfield}

\subsection{Derivation of general mean-field solution}

In this section, we study a curved neural network composed of $N$ neurons that stores $M$ patterns $\bm \xi^a=(\xi_1^a,\dots,\xi_N^a)$, as described by the deformed exponential family distribution given by
\begin{equation}
    p_\gamma(\bm x) = \exp\pr{-\varphi_\gamma}\Bigg[1+\gamma \beta \Bigg( H \sum_{a,i} \xi_i^a x_i + \frac{J}{N}\sum_{a,i<j} x_i  \xi_i^a \xi_j^a x_j\Bigg)\Bigg]_{+}^{1/\gamma},
\end{equation}
where $\varphi_\gamma$ is the normalising potential and $\gamma$ is the deformation parameter. 
In the following sections, we assume that parameters are scaled so that the content of the brackets $\brpos{\,}$ is always positive to avoid non-differentiable values.

We start the analysis by computing the value of $\exp\pr{\varphi}$ in the large $N$ limit, which can be done employing a delta integral substituting the value of $\frac{1}{N}\sum_i x_i$:
\begin{align}
    \exp\pr{\varphi_\gamma}&= \sum_{\bm x} \Bigg[1+\gamma \beta \Bigg( H \sum_{a,i} \xi_i^a x_i + \frac{J}{N}\sum_{a,i<j} x_i  \xi_i^a \xi_j^a x_j\Bigg)\Bigg]^{1/\gamma}_{+} \nonumber
    \\  &= \sum_{\bm x} \exp\pr{ \frac{1}{\gamma} \ln \pr{1+\gamma\beta \pr{H \sum_{a,i} \xi_i^a x_i + \sum_a \frac{J}{2N}\pr{ \bigg(\sum_{i}  \xi_i^a x_i \bigg)^2 -N}}}},
\end{align}
where the second equality uses $(\sum_i x_i)^2 - N=2\sum_{i<j}x_i x_j$. Additionally, by replacing $\frac{1}{N}\sum_i \xi_i^a x_i$ by a Dirac delta function under an integral, and then using the delta function's integral form $\delta\pr{x-a}=\frac{1}{2\pi} \int_{-\infty}^\infty e^{-\iu \zeta(x-a)}\,d\zeta$, the expression above can be re-written as
\begin{align}
    \exp\pr{\varphi_\gamma}&= \sum_{\bm x} \int d\bm m  \prod_a \delta\pr{m_a-\frac{1}{N}\sum_{i}\xi_i^a x_i} \exp\pr{ \frac{1}{\gamma} \ln \pr{1+\gamma N \beta \sum_a  \pr{ H m_a + \frac{J}{2}\pr{ m_a^2-\frac{1}{N}}}} } \nonumber
    \\    &= \frac{1}{(2\pi)^M} \int d\bm m d\bm{\hat  m} \sum_{\bm x} \exp\pr{ \frac{1}{\gamma} \ln \pr{1+\gamma N \beta \sum_a  \pr{ H m_a + \frac{J}{2}\pr{ m_a^2-\frac{1}{N}}}} - \sum_a \iu\hat m_a \bigg(m_a-\frac{1}{N}\sum_{i}\xi_i^a x_i\bigg)}. 
\end{align}

Let us now introduce a scaling rule for the deformation parameter $\gamma$ given by
\begin{equation}
    \gamma = \frac{\gamma'}{\beta N},
    \label{eq:scaled_gamma}
\end{equation}
where $\gamma'$ is a constant independent of $N$, which is motivated by subsequent results for the mean-field solution that suggest this relationship between $\gamma$ and $N$ in order to maintain scale-invariant properties. Then, the potential $\varphi_{\gamma}$ can be expressed in terms of $\gamma'$ as
\begin{align}
    \exp\pr{\varphi_{\gamma}} &= \frac{1}{(2\pi)^M} \int d\bm m d\bm{\hat  m} \exp\left( \frac{N \beta}{\gamma'} \ln \pr{1+\gamma' \sum_a  \pr{ H m_a + \frac{J}{2}\pr{ m_a^2-\frac{1}{N}}}} \right. \nonumber \\
    & \qquad \qquad \qquad \qquad \qquad \qquad
    \left. - \sum_a \iu\hat m_a m_a + \sum_i \ln\pr{2 \cosh\pr{ \frac{1}{N}\sum_a \xi_i^a \iu\hat m_a}}  \right).
\end{align}
Under this condition, the exponent in the equation above goes to infinity as $N\to\infty$. In this limit, the integral can be evaluated by the method of the steepest descent  (a.k.a. the saddle-point method), yielding
\begin{align}
    \exp\pr{\varphi_{\gamma}} &= \exp\left\{ \frac{N \beta}{\gamma'} \ln \pr{1+\gamma' \sum_a  \pr{ H m_a + \frac{J}{2}\pr{ m_a^2-\frac{1}{N}}}} \right. \nonumber \\
    & \qquad \qquad \qquad \qquad \qquad \qquad
    \left. - \sum_a \iu\hat m_a m_a + \sum_i \ln\pr{2 \cosh\pr{ \frac{1}{N}\sum_a \xi_i^a \iu\hat m_a}}  \right\}
\end{align}
where the content of the brackets set to the values of $\bm m, \bm{\hat m}$ that extremises (i.e. maximise or minimise) its content. To obtain their values we find the values that make the derivative of the expression inside the brackets equal to zero. 
By differentiating the exponent by $\hat m_a$, we find the saddle point must satisfy
\begin{equation}
    m_a = \frac{1}{N}\sum_i \xi_i^a \tanh \pr{\frac{1}{N} \sum_b \xi_i^b \iu\hat m_b}.
\end{equation}
Similarly, differentiating by $m_a$ yields
\begin{equation}
    \iu\hat m_a = \beta' N (H +J m_a),
\end{equation}
where we introduced the effective inverse temperature $\beta'$: 
\begin{equation}
    \beta' = \frac{ \beta }{1+\gamma' \sum_b \pr{Hm_b + \frac{J}{2} m_b^2}}.
\end{equation}
From these equations, we find the mean-field solution in the limit of large $N$:
\begin{equation}
    m_a = \frac{1}{N}\sum_i \xi_i^a \tanh \pr{\beta'\sum_b \xi_i^b (H+ Jm_b)},
    \label{eq-app:mf-Hopfield}
\end{equation}
which recovers the classical mean field solution at $\gamma'=0$. This solution confirms that $\gamma$ has to be scaled by the system size to maintain the scale-invariant properties.

The normalising potential in the large $N$ limit is obtained as
\begin{equation}\label{eq_app:neg_norm_potential}
    \varphi_{\gamma} =  \frac{\beta N}{\gamma'} \ln \frac{\beta'}{\beta} - \sum_a m_a  \beta' N (H+J m_a)  + \sum_i \ln\pr{2  \cosh\pr{ \beta'\sum_a \xi_i^a (H+J m_a)}}.
\end{equation}

\subsection{A single pattern: explosive phase transitions}

When embedded memory contains only a single pattern ($M=1$), the equations above result in
\begin{equation}
    \varphi_{\gamma} =   \frac{\beta N}{\gamma'} \ln \frac{\beta'}{\beta} -  N   \beta' (H+J m)m  + N \ln\pr{2 \cosh\pr{ \beta' (H+J m)}},
\end{equation}
with    
\begin{align}
    m &= \tanh\pr{\beta' (H+Jm)},
    \\ \beta' &= \frac{\beta}{1+\gamma' (Hm + \frac{J}{2} m^2)}.
    \label{eq:mean_field_m_simplified}
\end{align}
Under the limit of small $\gamma$ given by the scaling \eqref{eq:scaled_gamma}, the derivative of the normalisation potential $\varphi_\gamma$ w.r.t. $H$ yields the corresponding expected value, similarly to the exponential family distribution. Then, $\gamma'= 0$ yields the classical result. This can be verified by
\begin{align}
    \frac{\partial \varphi_{\gamma}}{\partial H} &=  -\frac{\beta N}{\gamma'} \frac{\partial \beta'}{\partial H}
    - N \beta' (H+Jm)\frac{\partial m}{\partial H}   - N \frac{\partial \beta'(H+Jm)}{\partial H} m 
    + N \frac{\partial \beta'(H+Jm)}{\partial H} m \nonumber
    \\ &=   - \frac{\beta N}{\gamma'} \frac{\partial \beta'}{\partial H}
    - N \beta' (H+Jm)\frac{\partial m}{\partial H}, 
\end{align}
where 
\begin{align}
    \frac{\partial \beta'}{\partial H} =  \frac{-\beta \gamma' \pr{ m + \frac{\partial m}{\partial H} (H + J m)} }{\pr{ 1+\gamma' (Hm + \frac{J}{2} m^2)}^2} = -\beta^{-1} \beta'^2 \gamma' \pr{m +   \frac{\partial m}{\partial H} (H + J m)},
    \label{eq:derivative_betaprime_by_H}
\end{align}
leading to 
\begin{align}
    \frac{\partial \varphi_{\gamma}}{\partial H} &= N\beta' \pr{m +   \frac{\partial m}{\partial H} (H + J m)} - N \beta' (H+Jm)\frac{\partial m}{\partial H} 
    \\&= N\beta'm.
\end{align}
The result recovers the classical relation, $\frac{\partial \varphi_{\gamma}}{\partial H} = \beta N m$ for the case $\gamma'=0$.

\subsubsection{Behaviour at criticality}

Now, we compute the critical exponents of the mean-field parameter for $H=0$. In the thermodynamic limit with $\gamma=\gamma'/(\beta N)$, one finds that
\begin{equation}
    m  =\tanh \pr{\frac{\beta J m}{1+\gamma' \frac{J}{2} m^2}}.
\end{equation}
Since $\tanh \pr{\frac{am}{1+bm}} = a m - (a^3/3 + ab) m^3 + O(m^4)$, by expanding the r.h.s. around $m=0$ up to the third order, one can find that
\begin{equation}
    m = \beta J m -  \frac{1}{6} (\beta J) \pr{ 2 (\beta J)^2  + 3 J \gamma'} m^3 + \mathcal{O}(m^4),
\end{equation}
which yields a trivial solution at $m=0$ and two non-trivial solutions given by
\begin{equation}
    m_{\pm} =\pm \sqrt{\frac{\beta J - 1}{ \frac{1}{6} \beta J \pr{ 2 (\beta J)^2  + 3 J \gamma'} } },
\end{equation}
which yields a mean-field universality class critical exponent `beta' (not to be confused with the inverse temperature) of $\frac{1}{2}$.

The magnetic susceptibility, $\chi \coloneqq \frac{\partial m}{\partial H}$, of the deformed Ising model can be calculated using \eqref{eq:mean_field_m_simplified}. Hence, we have
\begin{equation}
    \frac{\partial m}{\partial H} = (1-m^2) \left(\beta'\pr{1 +  J \frac{\partial m}{\partial H}} +  \frac{\partial \beta'}{\partial H} (H+Jm)\right).
\end{equation}
Using \eqref{eq:derivative_betaprime_by_H}, we obtain
\begin{align}
     \frac{\partial m}{\partial H} &= (1-m^2) \left(\beta'\pr{1 +  J \frac{\partial m}{\partial H}} -\beta^{-1} \beta'^2 \gamma' \pr{m +   \frac{\partial m}{\partial H} (H + J m)} (H+Jm)\right) \nonumber \\
     &= (1-m^2) \beta' \left( 1 +  J \frac{\partial m}{\partial H} -\frac{\beta'}{\beta}  \gamma' \pr{m +   \frac{\partial m}{\partial H} (H + J m)} (H+Jm)\right) \nonumber \\
    &= (1-m^2) \beta' \left( 1 -\frac{\beta'}{\beta}  \gamma' m (H + J m) +  \frac{\partial m}{\partial H} \pr{ J  -\frac{\beta'}{\beta} \gamma'  (H + J m)^2}\right).
\end{align}
Then, we obtain 
\begin{equation}
    \frac{\partial m}{\partial H} = \frac{ (1-m^2) \beta' \pr{1 -\frac{\beta'}{\beta}  \gamma' m (H + J m) } } 
    {1 - (1-m^2) \beta' \pr{ J  -\frac{\beta'}{\beta} \gamma'  (H + J m)^2} }.
    \label{eq:susceptibility_nonzero_gamma_prime}
\end{equation}

The susceptibility $\frac{d m}{d H}$ at $m=0$ is
\begin{align}
    \frac{\partial m}{\partial H} &= \frac{ \beta } 
    {1 -  \beta \pr{ J  -\frac{\beta'}{\beta} \gamma'  H^2} } \\
    &=  \frac{-\beta \beta_c}{\beta- \beta_c },
\end{align}
where the critical inverse temperature is given by $\beta_c = 1/(J  -\frac{\beta'}{\beta} \gamma'  H^2)$. Thus, the susceptibility results in the universality class `gamma' exponent of 1 (not to be confused with the deformation parameter) near the critical temperature. At $H=0$, $\beta_c = 1/J$.

Furthermore, at $\gamma' = 0$, we recover 
\begin{equation}
    \frac{\partial m}{\partial H} = \frac{ (1-m^2) \beta } 
    {1 - (1-m^2) \beta J }.
\end{equation}

\subsection{Two correlated patterns}
\label{app:two-patterns}

Here we study an exemplary case in which two patterns are embedded in the deformed associative network, with $\frac{1}{N}\sum_i \xi_i^1 \xi_i^2=C$. Thus, the fraction of terms for which $\xi_i^1 \xi_i^2=\pm 1$ is equal to $\frac{1\pm C}{2}$. We seek solutions for \eqref{eq-app:mf-Hopfield} and \eqref{eq:neg_norm_potential}, which can be further simplified for the case of two correlated patterns. We note that the content of the $\tanh$ and $\cosh$ terms can only take two values for the two patterns $\xi_i^a=\pm 1$. In the case of the $\tanh$ terms
\begin{align}
    \frac{1}{N}\sum_i \xi_i^a \tanh \pr{\beta'\sum_b \xi_i^b (H+ Jm_b)} =&  \frac{1}{N}\sum_i \tanh \pr{\beta'\sum_b \xi_i^a \xi_i^b (H+ Jm_b)} 
    \nonumber \\ =&  \frac{1+C}{2} \tanh\pr{ \beta'(2H + J (H+J (m_1+m_2))} +  \frac{1-C}{2} \tanh\pr{ \beta'J (m_1-m_2)}.
\end{align}
Hence, by replacing terms, one can find that
\begin{align}
    m_1 &=  \frac{1+C}{2} \tanh \pr{ \beta'  (2H+J m_1 + J m_2)} + \frac{1-C}{2} \tanh \pr{ \beta'  (J m_1 - J m_2)},
    \\  m_2 &= \frac{1+C}{2} \tanh \pr{ \beta'  (2H+J m_1 + J m_2)} - \frac{1-C}{2} \tanh \pr{ \beta'  (J m_1 - J m_2)}.
\end{align}
The normalising potential then becomes
\begin{align}
    \varphi_{\gamma} &=  \frac{\beta N}{\gamma'} \ln \frac{\beta'}{\beta} - \beta' N  m_1  (H+J m_1)  - \beta' N  m_2  (H+J m_2)
    \nonumber\\ &\quad\,  + \frac{1+C}{2}N  \ln\pr{2 \cosh\pr{ \beta'  (2H+J m_1 + J m_2)}} + \frac{1-C}{2}N  \ln\pr{2 \cosh\pr{ \beta'  (J m_1 - J m_2)}}.
\end{align}

\newpage

\section{Dynamical mean-field theory}
\label{app:dynamical_mft}

Let us now describe the statistics of temporal trajectories of the system. For this, let's consider the trajectory $\bm x_{0:T} = (\bm x_0,\dots,\bm x_T)$, whose probability can be computed as
\begin{equation}
    p_\gamma(\bm x_{0:T})= \prod_t p_\gamma(\bm x_t|\bm x_{t-1}),
\end{equation}
where the probability of the transition between $\bm x_{t-1}$ and $\bm x_t$ can be expressed as
\begin{equation}
    p_\gamma(\bm x_t|\bm x_{t-1}) = \frac{1}{N} \sum_i p_\gamma(x_{i,t}|\bm x_{t-1}) \prod_{j:j\neq i} \delta[x_{j,t},x_{j,t-1}],
\end{equation}
using the Kronecker delta, $\delta[x,y]$. For large system sizes, individual transitions (see \eqref{eq:glauber-approximation}) can be expressed as
\begin{align}
    p_\gamma(x_{i,t}|\bm x_{t-1})  &= \frac{\exp\pr{\beta' x_{i,t} h_{i,t}}}{2\cosh\pr{\beta' h_{i,t}}},
    \\h_{i,t} &=  \sum_a  \xi_i^a \bigg( H_a + \frac{1}{N}\sum_{j:j\neq i}  \xi_j^a x_{j,t-1}\bigg),
    \\ \beta_{t}' &= \frac{\beta}{1 + \gamma' \pr{\frac{1}{N}\sum_i x_{i,t-1}\sum_a \xi_i^a \pr{ H_a + 
    \frac{1}{2N}\sum_{j:j\neq i} \xi_j^a x_{j,t-1} }}}.
\end{align}
As before, the above derivation assumes that the content of the $\brpos{\,}$ operator in the definition of the deformed exponential family is non-negative.

Using the integral form of the Kronecker delta function, the above transition probability can be rewritten as
\begin{align}
    p_\gamma(\bm x_t|\bm x_{t-1})
    &= \frac{1}{N}\sum_i p_\gamma(x_{i,t}|\bm x_{t-1}) \prod_{j:j\neq i} \delta[1,x_{j,t}x_{j,t-1}]
    \nonumber \\ &= \frac{1}{N(2\pi)^{N-1} } \sum_i \int_{\bm 0}^{\bm 2 \pi} d\bm\phi_t \exp\vast( \beta'_{t} x_{i,t} h_{i,t} - \ln\pr{2\cosh\pr{\beta_t' h_{i,t}}}
    + \sum_{j:j\neq i }\iu \phi_{j,t} (1-x_{j,t}x_{j,t-1}) \vast).
\end{align} 
Let $k_t\in (1,\dots,N)$ be a uniform independent random variable. Namely, at each time step $t$, the index $k_t$ is drawn independently and uniformly from the set $\{1,2,\ldots,N\}$. Then, the sequence $\{k_t\}_{t=1}^{T}$ constitutes auxiliary variables to keep track of which spin is being updated at each time step. Using $k_t$, the average over the spin in the equation above can be replaced by an average over the uniform $k_t$: $  \frac{1}{N}\sum_{i=1}^N \;\rightarrow\; \sum_{k_t=1}^N\frac{1}{N}$. Using this, the probability of the trajectory $\bm x_{0:t}$ can be rewritten as
\begin{align}
    p_\gamma(\bm x_{0:T}) &= \frac{1}{N^T(2\pi)^{(N-1)T}}  \sum_{\bm k} \int_{\bm 0}^{\bm 2 \pi} d\bm\phi \exp\vast( \sum_{i,t} \vast( \pr{\beta'_{t} x_{i,t} h_{i,t} -\ln\pr{2\cosh\pr{\beta_t' h_{i,t}}}}\delta[i,k_t]
    \nonumber\\  & \qquad \qquad \qquad \qquad \qquad \qquad\qquad- \sum_{j:j\neq i}\iu \phi_{j,t} (1-x_{j,t}x_{j,t-1}) (1-\delta[j,k_t])\vast)\vast)
    \nonumber \\  &= \frac{1}{N^T(2\pi)^{(N-1)T}}  \sum_{\bm k} \int_{\bm 0}^{\bm 2 \pi} d\bm\phi \exp\vast( \sum_{i,t} \vast( \pr{\beta'_{t} x_{i,t} h_{i,t} -\ln\pr{2\cosh\pr{\beta_t' h_{i,t}}}}\delta[i,k_t]
    \nonumber\\  & \qquad \qquad \qquad \qquad \qquad \qquad\qquad - (N-1)\iu \phi_{i,t} (1-x_{i,t}x_{i,t-1}) (1-\delta[i,k_t])\vast)\vast)
    \nonumber\\ =& \lim_{\lambda\to\infty} \frac{1}{N^T}  \sum_{\bm k}  \exp\vast( \sum_{i,t} \vast( \pr{\beta'_{t} x_{i,t} h_{i,t} -\ln\pr{2\cosh\pr{\beta_t' h_{i,t}}}}\delta[i,k_t]
   - \lambda (N-1) (1-x_{i,t}x_{i,t-1}) (1-\delta[i,k_t])\vast)\vast),
   \label{eq:prob_trajectory_2}
\end{align}
where the second equality switches $i,j$ indices and the third equality can be justified by the $\Gamma$-convergence of the log probability functionals
\begin{equation}
      \limsup_{\lambda \to \infty} \left( -\lambda C  (1 - x y)(1-\delta)\right) \ge \ln \frac{1}{2\pi}\int_0^{2\pi} d\phi \, \exp\left( -i \phi  \, C (1 - x y)(1-\delta) \right),
\end{equation}
which substitutes the oscillatory delta integral by a soft exponential penalty with a large parameter $\lambda$, ensuring that the minimizers of the soft-penalized functional converge to those of the original constrained system.

As (with $\delta\in[0,1]$) we have $\lim_{\lambda\to\infty} \ln \cosh(k\pm\lambda(1-\delta))  =\lim_{\lambda\to\infty}( \ln \cosh(k)\delta + \lambda (1-\delta) )$, we can simplify the equation above as follows:
\begin{align}
    p_\gamma(\bm x_{0:T}) &= \lim_{\lambda\to\infty} \frac{1}{N^T}  \sum_{\bm k}  \exp\vast( \sum_{i,t}\pr{\beta'_{t} x_{i,t} h_{i,t}^{\lambda, k_t}-\ln\pr{2\cosh\pr{\beta_t' h_{i,t}^{\lambda, k_t}}}}\vast),
   \\ h_{i,t}^{\lambda, k_t}=& \sum_a  \xi_i^a \pr{ H_a + \frac{1}{N}\sum_{j:j\neq i}  \xi_j^a x_{j,t-1}} + \beta_t'^{-1} \lambda(N-1) x_{i,t-1}  (1-\delta[i,k_t]) .
\end{align}
This operation absorbs the second term in \eqref{eq:prob_trajectory_2} related to ``the spin $i$ not chosen'' ($\delta[i,k_t]=0$) into the effective field $ h_{i,t}^{\lambda, k_t}$, ensuring the strong coupling of the current state $x_{i,t}$ with the previous state $x_{i,t-1}$ using large $\lambda$. One can verify that it recovers \eqref{eq:prob_trajectory_2} by noting that if $\delta[i,k_t]=1$ (spin $i$ chosen), then $h_{i,t}^{\lambda, k_t}=h_{i,t}$, recovering the first term in the exponent of \eqref{eq:prob_trajectory_2}, and that if $\delta[i,k_t]=1$ (spin $i$ not chosen), we have $h_{i,t}^{\lambda, k_t} \sim \beta_t'^{-1} \lambda(N-1) x_{i,t-1} $ for large $\lambda$, which yields
\begin{align}
    \beta'_{t} x_{i,t} h_{i,t}^{\lambda, k_t} - \ln\pr{2\cosh\pr{\beta_t' h_{i,t}^{\lambda, k_t}}} 
    &\sim \lambda(N-1) x_{i,t} x_{i,t-1}
    - | \lambda(N-1) x_{i,t-1} | = \lambda(N-1)(1 - x_{i} x_{i,t-1}),
\end{align}
recovering the second term in \eqref{eq:prob_trajectory_2}.

In equilibrium systems, the partition function retrieves their statistical moments. A nonequilibrium equivalent function is a generating functional or dynamical partition function \cite{coolen2001statisticalII}.
Let us now define the generating functional
\begin{equation}
    Z(\bm g) = \sum_{\bm x}   \exp\pr{ \sum_{i,t} g_{i,t} x_{i,t}} p_\gamma(\bm x_{0:T}),
\end{equation}
such that the following relationship is satisfied:
\begin{equation}
    \frac{d Z(\bm 0)}{d g_{i,t}} = \ang{x_{i,t}}.
\end{equation}
Then, one can find an analytical expression for the functional by introducing delta integrals. Defining
\begin{align}
    \tilde h_{i,t}^{\lambda, k_t}=& \sum_a \xi_i^a (H_a + m_{a,t-1}) +  \beta_t'^{-1} (N-1)\lambda x_{i,t-1}  (1-\delta[i,k_t]) ,  \label{eq:tilde-h-k}
    \\\tilde \beta_{t}' &= \frac{\beta}{1 + \gamma'\sum_a  \pr{H_a m_{a,t-1} +\frac{1}{2} m_{a,t-1}^2}},
     \label{eq:tilde-beta}
\end{align}
we obtain
\begin{align}
    Z(\bm g) &= \lim_{\lambda\to\infty}\frac{1}{N^T}  \sum_{\bm x} \sum_{\bm k} \exp\vast( \sum_{i,t}  \pr{x_{i,t} (g_{i,t}+\tilde\beta'_{t}  h_{i,t}^{\lambda, k_t}) - \ln\pr{2\cosh\pr{\tilde\beta_t' h_{i,t}^{\lambda, k_t}}}}\vast) \nonumber\\
    &= \lim_{\lambda\to\infty}\frac{1}{N^T(2\pi)^{MT}}  \sum_{\bm x} \int  d\bm m d\bm{\hat m}  \sum_{\bm k}\exp\vast( \sum_{i,t}\pr{ x_{i,t}(g_{i,t}+\tilde{\beta}'_{t}  \tilde h_{i,t}^{\lambda, k_t}) - \ln\pr{2\cosh\pr{\tilde{\beta}_t' \tilde h_{i,t}^{\lambda, k_t}}}}\nonumber\\
    &\quad\; - \sum_{a,t} \iu \hat m_{a,t} \pr{m_{a,t} - \frac{1}{N}\sum_i \xi_i^a x_{i,t} }\vast)
    \nonumber\\ &= \lim_{\lambda\to\infty}\frac{1}{(2\pi)^{MT}}  \int d\bm m d\bm{\hat m}  \exp\vast(   - \sum_{a,t} \iu \hat m_{a,t} m_{a,t} 
     + \ln \pr{\frac{1}{N^T}\sum_{\bm x, \bm k} e^{L^{\bm k}}}  \vast),
\end{align}
where $L^{\bm k}$ is given as
\begin{align}
    L^{\bm k} = \sum_{i,t} x_{i,t} \pr{g_{i,t}+\tilde{\beta}'_{t}  \tilde h_{i,t}^{\lambda, k_t} + \sum_a \xi_i^a \iu \hat m_{a,t}} -\sum_{i,t} \ln\pr{2\cosh\pr{\tilde{\beta}_t' \tilde h_{i,t}^{\lambda, k_t}}}.
\end{align}

One can solve the mean-field equations via steepest descent, obtaining
\begin{align}
    m_{a,t} =& \frac{1}{N}\sum_i \xi_i^a \ang{x_{i,t}}_{L^{\bm k}},
    \\ \iu\hat m_{a,t} =& \sum_a \xi_i^a \ang{  \pr{x_{i,t} - \tanh\pr{{\tilde{\beta}_t' \tilde h_{i,t}^{\lambda, k_t}}}} }_{L^{\bm k}}
\end{align}
with
\begin{align}
    \ang{f(\bm x)}_{L^{\bm k}} = \lim_{\lambda\to\infty}\frac{\frac{1}{N^T}\sum_{\bm x, \bm k} f(\bm x)e^{L^{\bm k}}}{\frac{1}{N^T}\sum_{\bm x, \bm k} e^{L^{\bm k}}}.
\end{align}
For $\bm g = \bm 0$, we obtain $\iu\hat m_{a,t}=0$ and $\frac{1}{N^T}\sum_{\bm x, \bm k} e^{L^{\bm k}}=1$.

Let us also define as
\begin{align}
    \ang{f(\bm x)}_{L} =& \frac{\sum_{\bm x} f(\bm x)e^{L}}{\sum_{\bm x} e^{L}},
\end{align}
where $L$ is defined 
\begin{align}
    L =& \sum_{i,t} x_{i,t} \pr{g_{i,t}+\tilde{\beta}'_{t}  \tilde h_{i,t} + \sum_a \xi_i^a \iu \hat m_{a,t}} -\sum_{i,t} \ln\pr{2\cosh\pr{\tilde{\beta}_t' \tilde h_{i,t}}},
\end{align}
using 
\begin{align}
    \tilde h_{i,t} &= \sum_a \xi_i^a (H_a + m_{a,t-1}).
    \label{eq:tilde-h} 
\end{align}
One can relate $L$ with $L^{\bm k}$ via the following equation, 
\begin{align}
    \lim_{\lambda\to\infty}e^{L^{\bm k}} =& e^{L} \prod_{i,t} (1-\delta[i, k_{t}]) \delta[x_{i,t}x_{i,t-1}]
\end{align}
forcing spins not to change sign when $\delta[i, k_{i,t}]=0$. Note that, when $k_{t}\neq i$, spin $x_{i,t}$ takes the value of the spin at the previous step $x_{i,t-1}$. 

The mean field behaviour is  recovered for $\bm g = \bm 0$, which we will assume from now on.
This results in
\begin{align}
     \ang{x_{i,t}}_{L^{\bm k}} =& \lim_{\lambda\to\infty} \frac{1}{N^T} \sum_{\bm x_t \bm k} x_{i,t} e^{L^{\bm k}} 
     \nonumber\\ =& \lim_{\lambda\to\infty} \pr{ \frac{1}{N^{t-1}}\sum_{\bm x_t \bm k: k_t\neq i} x_{i,t} e^{L^{\bm k}} + \frac{1}{N^{t-1}}\sum_{\bm x_t \bm k: k_t=i} x_{i,t} e^{L^{\bm k}} }
     \nonumber\\ =& \bigg(1-\frac{1}{N}\bigg)\ang{x_{i,t-1}}_{L^{\bm k}} + \frac{1}{N} \ang{x_{i,t}}_L
\end{align}
because of the effect of the $\delta$ function in \eqref{eq:tilde-h-k}.

We also find that
\begin{align}
     \ang{x_{i,t}}_{L} = \tanh\pr{\tilde\beta'_t\tilde h_{i,t}}.
\end{align}
This results in
\begin{align}
    m_{a,t} =&\frac{1}{N} \sum_i\xi_i^a \ang{x_{i,t}}_{L^{\bm k}}  = m_{a,t-1}\pr{1-\frac{1}{N}} +\frac{1}{N^2}\sum_i \xi_i^a \tanh\pr{\tilde\beta'_t \tilde h_{i,t}}.
\end{align}
The expression above can be rearranged in the form similar to a differential equation
\begin{align}
   \frac{m_{a,t} - m_{a,t-1}}{N^{-1}}  =& = -m_{a,t-1} +\frac{1}{N}\sum_i \xi_i^a \tanh\pr{\tilde{\beta}_t'\sum_b \xi_i^b(H_b+ m_{b,t-1})}.
\end{align}
Under large $N$ and for an adequate time re-scaling, this leads to the following differential equation:
\begin{align}
    \dot m_{a} &= -m_{a} + \frac{1}{N}\sum_i \xi_i^a \tanh\pr{\beta'\sum_b \xi_i^b(H_b+ m_{b})},
    \\ \beta' &= \frac{\beta}{1 + \gamma'\sum_a \pr{H_a m_a +\frac{1}{2} m_{a}^2}}.
    \label{eq-app:dynamical-mean-field}
\end{align}

\newpage
\section{Replica analysis near saturation}
\label{app:near_saturation}

Here we analyse a curved neural network with an extensive number of patterns, $M=\alpha N$ in \eqref{eq:deformed-Hopfield-net}. The model involves integrals over a large number of variables, making the steepest descent method inapplicable. Instead, we adopt the approach reported in Ref.~\cite{amit1985storing}, and average the free energy over the distribution of patterns using the replica trick. 

For $Z=\exp(\varphi_\gamma)$, the replica trick is applied as follows:
\begin{equation}
    \llangle \varphi_\gamma \rrangle =\llangle\ln Z\rrangle = \lim_{n\to 0} \frac{1}{n}(\llangle Z^n \rrangle - 1),
\end{equation}
which can be equivalently written as
\begin{equation}
    \llangle \ln Z \rrangle = \lim_{n\to 0} \frac{1}{n}\ln\llangle Z^n \rrangle
\end{equation}
with $ \llangle f(\bm x) \rrangle = 2^{-MN}\sum_{\bm \xi} f(\bm x) $ being the configurational average over different combinations of the systems' parameters.

\subsection{General derivation}

To calculate the encoding of patterns, we introduce $\{\bm\xi_{a} \}$ with $a=1,...,M$ where the first $l$ patterns are given --- called `nominated' patterns --- and we average over the $M-l$ rest. Again, assuming as in Supplementary Note~\ref{app:hopfield} that the content of the $\brpos{\,}$ is positive, we calculate
\begin{equation}
    \llangle Z^n \rrangle = \frac{1}{2^{N(M-l)}}\sum_{\bm\xi^{a>l}}\sum_{\bm x} \exp\pr{\frac{1}{\gamma}\sum_u \ln\pr{1+\gamma \beta \pr{\sum_{b\leq l} H_b \frac{1}{N}\sum_{i} x_i^u \xi_i^b + \frac{J}{N}\sum_a \sum_{i<j}  x_i^u  \xi_i^a \xi_j^a x_j^u} }}.
\end{equation}

We want to compute the configurational average of a network with $M$ memories with $N,M\to\infty$ and $M/N=\alpha$, introducing a pair of delta integrals
\begin{align}
    \llangle Z^n \rrangle&=  \frac{1}{2^{N(M-l)}\pr{2\pi}^{(l+1)n}} \int d\bm m d\bm{\hat m} d\bm \mu d\bm{\hat \mu}\sum_{\bm\xi^{a>l}} \sum_{\bm x} \exp\vast(
    -\sum_{u,b\leq l}\iu \hat m_b^u \bigg(m_b^u - \frac{1}{N} \sum_i x_i^u \xi_i^b\bigg)
    \nonumber\\ &\quad\, 
    - \sum_u \iu \hat \mu_u \Bigg(\mu_u -\frac{1}{2}\sum_{a>l}  \pr{\frac{1}{\sqrt{N}}\sum_{i}  x_i^u  \xi_i^a}^2\Bigg)
    \nonumber\\ &\quad\, 
    \frac{1}{\gamma} \sum_u \ln\pr{1+\gamma \beta \pr{ \sum_{b\leq l} \pr{N H_b m_b^u + \frac{JN}{2} \pr{m_b^u}^2 }+ J \mu_u- N\frac{J\alpha}{2}}}
    \vast),
\end{align}
where the $ N\frac{J\alpha}{2}$ comes from substracting the diagonal. This leads to
\begin{align}
    \llangle Z^n \rrangle&= \frac{1}{\pr{2\pi}^{(l+1)n}} \int d\bm m d\bm{\hat m}  d\bm \mu d\bm{\hat \mu}\sum_{\bm x} \exp\vast(
    \frac{1}{\gamma} \sum_u \ln\pr{1+\gamma \beta \pr{ \sum_{b\leq l} \pr{N H_b m_b^u + \frac{JN}{2} \pr{m_b^u}^2} + J \mu_u- N\frac{J\alpha}{2}}}
    \nonumber\\ & -\sum_{u,b\leq l}\iu \hat m_b^u \bigg(m_b^u - \frac{1}{N} \sum_i x_i^u \xi_i^b\bigg) - \sum_u \iu \hat \mu_u \mu_u + \ln \left(\frac{1}{2^{N(M-l)}}\sum_{\bm\xi^{a>l}}\exp\Bigg( \sum_u \frac{1}{2}\sum_{a>l}  \pr{\sqrt{\frac{\iu \hat \mu_u}{N}}\sum_{i}  x_i^u  \xi_i^a}^2 \Bigg)\right)\vast).
\end{align}
To compute the last term, we can integrate over disorder by factorising over patterns $a$ and introducing a Gaussian integral ($\int Dz \exp\pr{az} = \exp\pr{a^2/2} $) to obtain
\begin{align}
    \frac{1}{2^{N(M-l)}}\sum_{\bm\xi^{a>l}} \exp\vast(  \sum_{a>l} \sum_u \pr{  \frac{1}{2}\sqrt{\frac{\iu \hat \mu_u}{N}} \sum_{i}  x_i^u  \xi_i^a}^2\vast)&=  \frac{1}{2^{N(M-l)}}  \prod_{a> l} \sum_{\bm\xi^a} \exp \Bigg( \sum_u \frac{1}{2} \pr{ \sqrt{\frac{\iu \hat \mu_u}{N}}\sum_{i}  x_i^u  \xi_i^a}^2\Bigg)
    \nonumber\\ &= \prod_{a> l}  \frac{1}{2^{N(M-l)}} \sum_{\bm\xi^a} \int D\bm z \exp\Bigg(   \sum_u z_u \sqrt{\frac{\iu \hat \mu_u}{N}}\sum_{i}  x_i^u  \xi_i^a\Bigg)
    \nonumber\\ &=  \frac{1}{2^{N(M-l)}} \pr{  \int D\bm z \exp \Bigg( \sum_i \ln 2\cosh\pr{\sum_u \sqrt{\frac{\iu \hat \mu_u}{N}} z_u x_i^u }\Bigg)}^{M-l}
    \nonumber\\ &= \pr{  \int D\bm z \exp \Bigg( \sum_i \ln \cosh\pr{\sum_u \sqrt{\frac{\iu \hat \mu_u}{N}} z_u x_i^u }\Bigg)}^{M-l}
    \nonumber\\ &= \pr{ \int D\bm z \exp \Bigg(  \frac{1}{2}\sum_{u,v} z_u z_v \sqrt{  \iu \hat \mu_u \iu \hat \mu_v}\frac{1}{N} \sum_i x_i^u x_i^v \Bigg)}^{M-l},
\end{align}
where the $\ln \cosh$ term was approximated assuming a large $N$ ($\ln \cosh x \approx \ln (1 + x^2/2) \approx x^2/2$ for $x\ll1$). By introducing an additional delta integral for order parameters $q_{uv}$ (assuming $q_{uu}=1$) and applying $\exp \ln$, one can re-express the last term (assuming $M-l\approx N\alpha$ near saturation) as
\begin{align}
     & \exp\pr{ N\alpha \ln \int D\bm z \exp\vast(  \frac{1}{2}\sum_{u,v} z_u z_v \sqrt{  \iu \hat \mu_u \iu \hat \mu_v}\frac{1}{N} \sum_i x_i^u x_i^v \vast)}
     \nonumber\\ &= \frac{1}{(2\pi)^{\frac{n(n-1)}{2}}} \int d\bm q d\bm{\hat q}  \exp\vast(N\alpha \ln \int D\bm z \exp\vast( \frac{1}{2} \sum_{u,v} z_u z_v \sqrt{  \iu \hat \mu_u \iu \hat \mu_v}q_{uv} \vast)
    - \sum_{u<v} \iu \hat q_{uv} \pr{q_{uv} - \frac{1}{N} \sum_i x_i^u x_i^v}\vast)
    \nonumber\\ &= \frac{1}{(2\pi)^{\frac{n(n-1)}{2}}} \int d\bm q d\bm{\hat q}\exp\vast(  -\frac{1}{2}N\alpha\ln \abs{\Lambda} - \sum_{u<v} \iu \hat q_{uv} \pr{q_{uv} - \frac{1}{N} \sum_i x_i^u x_i^v}\vast) ,
\end{align}
where $\Lambda_{uv} = \delta_{uv} (1-\sqrt{\iu\hat\mu_u \iu\hat\mu_v} q_{uv}) + (1- \delta_{uv} ) \pr{-\sqrt{\iu\hat\mu_u \iu\hat\mu_v} q_{uv} } = \delta_{uv} - \sqrt{\iu\hat\mu_u \iu\hat\mu_v}q_{uv}$. Then, the configurational average is found to be
\begin{align}
    \llangle Z^n \rrangle &= \frac{1}{\pr{2\pi}^{(l+1)n+\frac{n(n-1)}{2}}} \int d \bm \pi
    \sum_{\bm x} \exp\vast(
    \frac{1}{\gamma} \sum_u \ln\pr{1+\gamma \beta \pr{ \sum_{b\leq l} \pr{N H_b m_b^u + \frac{JN}{2} \pr{m_b^u}^2} + J \mu_u- N\frac{J\alpha}{2}}}
    \nonumber\\ &\quad\, -\sum_{u,b\leq l}\iu \hat m_b^u \bigg(m_b^u - \frac{1}{N} \sum_i x_i^u \xi_i^b \bigg) - \sum_u \iu \hat \mu_u \mu_u -\frac{1}{2} N\alpha \ln\abs{\bm \Lambda} - \sum_{u<v} \iu \hat q_{uv} \bigg( q_{uv} - \frac{1}{N} \sum_i x_i^u x_i^v \bigg)\vast)
    \nonumber \\ &= \frac{1}{\pr{2\pi}^{(l+1)n+\frac{n(n-1)}{2}}}\int d \bm \pi
   \exp\vast(
    \frac{1}{\gamma} \sum_u \ln\pr{1+\gamma \beta \pr{ \sum_{b\leq l} \pr{N H_b m_b^u + \frac{JN}{2} \pr{m_b^u}^2} + J \mu_u- N\frac{J\alpha}{2}}}
    \nonumber\\ &\quad\, -\sum_{u,b\leq l}\iu \hat m_b^u m_b^u  - \sum_u \iu \hat \mu_u \mu_u -\frac{1}{2} N\alpha \ln\abs{\bm \Lambda} -\sum_{u<v} \iu \hat q_{uv}q_{uv}+ \ln \sum_{\bm x} \exp L \vast),
\end{align}
where $d \bm \pi \coloneqq d\bm m d\bm{\hat m}  d\bm \mu d\bm{\hat \mu} d\bm q d\bm{\hat q}$ has been adopted for readability, with
\begin{equation}\label{eq-app:L_function}
    L = \sum_{u,b\leq l} \iu \hat m_b^u \frac{1}{N} \sum_i x_i^u \xi_i^b + \sum_{u<v} \iu\hat q_{uv} \frac{1}{N}\sum_i x_i^u x_i^v
\end{equation}
carrying all remaining $x_i$ dependent terms to be summed. The saddle-node solution is given by
\begin{align}\label{eq-app:saddle_node_sol}
    \begin{aligned}
    \iu\hat m_a^u &= N\beta'_u (H_a +  J m_a^u),
    \\ \iu\hat\mu_{u} &= \beta'_u J ,
    \\  m_a^u &= \frac{1}{N} \sum_i \xi_i^a \ang{x_i^u }_{L} \coloneqq \frac{1}{N} \sum_i \xi_i^a \frac{\sum_{\bm x} x_i^u\exp L }{\sum_{\bm x}  \exp L},
    \\  q_{uv} &= \frac{1}{N} \sum_i \ang{x_i^u x_i^v}_{L} = \frac{1}{N} \sum_i  \frac{\sum_{\bm x} x_i^u x_i^v \exp L }{\sum_{\bm x}  \exp L},
    \\ \iu \hat q_{uv} &=N\alpha \sqrt{\iu\hat\mu_{u}\iu\hat\mu_{v}} \ang{z_u z_v}_*  = N\alpha J^2\beta'_u \beta'_v r_{uv},
    \\\mu_{u} &=   \frac{1}{2} N\alpha \sum_{v} \sqrt{\frac{\beta'_v }{\beta'_u }} q_{uv}  \ang{z_u z_v}_*   =  \frac{1}{2} N\alpha J \sum_{v} \sqrt{\beta'_u\beta'_v} q_{uv}  r_{uv},
    \\ \beta_u' &= \frac{\beta}{1+\gamma N \beta  \frac{1}{2} \pr{\sum_{b\leq l} (2H_b +  J m_b^u) m_b^u+ \alpha J^2  \sum_{v} \sqrt{\beta'_u \beta'_v}  q_{uv}  r_{uv} - J\alpha }}, 
    \\ r_{uv} &= \frac{1}{J\sqrt{\beta'_u\beta'_v}}\ang{z_u z_v}_*   \coloneqq \frac{1}{J\sqrt{\beta'_u\beta'_v}} \frac{\int D\bm z z_u z_v \exp\pr{\frac{1}{2} \sum_{u,v} z_u z_v  \pr{ \sqrt{ \beta'_u \beta'_v }J q_{uv}} }}{\int D\bm z \exp\pr{\frac{1}{2} \sum_{u,v} z_u z_v  \pr{ \sqrt{ \beta'_u \beta'_v }J q_{uv}} }},
    \end{aligned}
\end{align}
where the operator $\ang{f(\bm x)}_*$ defined above coincides with regular averages once integration is performed. This results in
\begin{align}
    \ln \llangle Z^n \rrangle&= 
    \frac{1}{\gamma} \sum_u \ln\pr{1+\gamma \beta \pr{ \sum_{b\leq l} \pr{N H_b m_b^u + \frac{JN}{2} \pr{m_b^u}^2} + J \mu_u- N\frac{J\alpha}{2}}}
    \nonumber\\ &\quad\, -\sum_{u,b\leq l} N\beta'_u (H_b + J m_b^u) m_b^u  - \sum_u \beta'_u J  \mu_u -\frac{1}{2} N\alpha \ln \abs{\bm\Lambda} - \sum_{u<v} N\alpha \beta'_u \beta'_v J^2 r_{uv} q_{uv}
    \nonumber\\ &\quad\, + \ln \sum_{\bm x} \exp L
\end{align}
with $L$ being given (due to \eqref{eq-app:L_function}) by
\begin{equation}
    L = \sum_{u,b\leq l} \beta'_u (H_b+  J m_b^u)\sum_i x_i^u \xi_i^b +  J^2\alpha\sum_{u<v}{ \beta'_u \beta'_v} r_{uv}\sum_i x_i^u x_i^v.
\end{equation}

\subsection{Replica symmetry}
\label{app:Replica_Sym}

The replica symmetry ansatz allows us to simplify order parameters $m_b^u ,q_{uv}, r_{uv}$ (for $u\neq v$), and $r_{uu}$ to homogeneous values $m_b, q, r$, and $R$ (note that $q_{uu}=1$). Assuming a normalised curvature parameter $\gamma=\frac{\gamma'}{N\beta}$, we obtain
\begin{align}
    \frac{1}{n}\ln \llangle Z^n \rrangle&= 
      \frac{N\beta}{\gamma'} \ln\pr{1+\gamma' \pr{ \sum_{b\leq l} \pr{H_b m_b+ \frac{J}{2} \pr{m_b}^2} + \frac{J}{N}\mu- \frac{J\alpha}{2}}}
    \nonumber\\ &\quad\; -\sum_{b\leq l} N \beta'_u (H_b + J m_b) m_b  -  \beta' J  \mu -\frac{1}{2n} N\alpha \ln\abs{\bm  \Lambda} 
    \nonumber\\ &\quad\;
    - \frac{1}{2}(n-1) N\alpha  \beta'^2 J^2 r q  
    + \frac{1}{n}\ln \sum_{\bm x} \exp L, 
\end{align}
where $ J\beta' \mu=\frac{1}{2} N\alpha (J \beta')^2 (R+(n-1)qr)$ and
\begin{equation}
    L = \beta'\sum_{u,b\leq l}(H_b + J m_b)  \sum_i x_i^u \xi_i^b +  \sum_{u<v}  \beta'^2 J^2 \alpha r \sum_i x_i^u x_i^v.
\end{equation}
We obtain
\begin{align} \label{eq_app:RS_Lfunction}
    \ln\sum_{\bm x} \exp L &= \ln\sum_{\bm x}  \exp\pr{ \sum_{u,b\leq l} \beta' (H_b +J m_b) \sum_i \xi_i^b  x_i^u + \pr{\beta' J}^2 \alpha r \sum_i \sum_{u<v} x_i^u  x_i^v } \nonumber
    \\ &= \ln\prod_i \sum_{\bm x_i} \exp\pr{   \beta' \sum_{u,b\leq l}(H_b +J m_b) \xi_i^b x_i^u  + \frac{1}{2} \pr{ \beta' J \sqrt{\alpha r} \sum_{u} x_i^u }^2 - \frac{1}{2} n (\beta' J)^2 \alpha r } \nonumber
    \\ &= \ln\prod_i \int Dz \sum_{\bm x_i} \exp\pr{\beta' \sum_{u,b\leq l} (H_b +J m_b) \xi_i^b x_i^u +  \beta' J \sqrt{\alpha r} z \sum_{u} x_i^u  -  \frac{1}{2} n (\beta' J)^2  \alpha r } \nonumber
    \\ &= \sum_i \ln \int Dz \exp\pr{ n \ln\pr{2 \cosh\pr{ \beta'  \sum_{b\leq l} (H_b +J m_b) \xi_i^b + \beta' J \sqrt{\alpha r} z  }}-  \frac{1}{2} n (\beta' J)^2 \alpha r} \nonumber
    \\ &= \sum_i n \int Dz \ln\pr{   2 \cosh\pr{ \beta' \sum_{b\leq l} (H_b +J m_b) \xi_i^b +  \beta' J \sqrt{\alpha r} z } }- \frac{1}{2} nN \pr{\beta' J}^2 \alpha r,
\end{align}
where in the last step we assum a small value of $n$ (as we will apply later the limit $n\to 0$).

\paragraph*{Overlaps corresponding to non-nominated patterns}

To apply the replica symmetry argument near the $n\to 0$ limit, we know from \eqref{eq-app:saddle_node_sol} that the covariance matrix of $\beta'J r_{uv}$ corresponds to the inverse of $\bm\Lambda$ (with $\Lambda_{u v} = \delta_{uv} - \beta' J (\delta_{uv}+(1-\delta_{uv})q ) = \delta_{uv}(1 - \beta' J (1-q)) - \beta' J q$). Under the replica symmetry assumption, the inverse is given  via the Sherman-Morrison formula,
\begin{equation}
    \Lambda^{-1}_{uv} = \frac{1}{1-\beta' J(1-q)} \pr{ \delta_{uv} + \frac{\beta' J q}{1-\beta' J(1-q) - n\beta' J q} } ,
\end{equation}
evaluating the limit $n=0$,
\begin{equation}
    \beta' J r_{uv} = \delta_{uv}\frac{1- \beta' J (1-2 q)}{(1 - \beta'J (1 - q))^2} +  (1-\delta_{uv})\frac{ \beta' J q}{(1 -\beta'J  (1 - q))^2}. 
\end{equation}
Identifying $R$ and $r$ as the respective diagonal and off diagonal parts of $r_{uv}$, we can determine
\begin{equation}
    \mu \xrightarrow{n \to 0} \frac{1}{2} N  \alpha \beta' J (R - qr)\,. 
\end{equation}
This leads to an effective inverse temperature,
\begin{equation}\label{eq-app:inv_temp_RS}
    \beta' = \frac{\beta}{1+\gamma'  \frac{1}{2} \pr{\sum_{b\leq l} (2 H_b +  J m_b) m_b + \alpha  J\pr{\frac{ 1 - \beta' J (1-q)^2}{(1 -\beta'J(1 - q))^2} -1}}}\,.
\end{equation}

\paragraph*{Replica symmetric solution}
We have
\begin{align}
    \frac{1}{n}\ln \llangle Z^n \rrangle&= 
      \frac{N\beta}{\gamma'} \ln\pr{1+\gamma' \pr{ \sum_{b\leq l} \pr{H_b m_b+ \frac{J}{2} \pr{m_b}^2} + \frac{J}{N} \mu- \frac{J\alpha}{2}}}
    \nonumber\\ & - N \beta' \sum_{b\leq l}(H_b +  J m_b) m_b  -  \beta' J  \mu -\frac{1}{2n} N\alpha \ln \abs{\Lambda}
    - \frac{1}{2} N\alpha (n-1) (J \beta')^2 r q  
    \nonumber \\& + \sum_i \int Dz \ln\pr{2 \cosh\pr{ \beta'  \sum_{b\leq l} (H_b +J m_b) \xi_i^b +  \beta' J \sqrt{\alpha r} z }}-  \frac{1}{2}N \pr{\beta' J}^2 \alpha r .
    \label{eq-app:RS_solution_Z}
\end{align}
Near the limit of $n\to0$, we can approximate
\begin{align}
    \ln \abs{\Lambda} &= \ln\pr{1-\beta'J(1-q)-\beta'J qn} + (n-1)\ln\pr{1-\beta'J(1-q)}
    \nonumber \\
    &= n \pr{ \ln\pr{1-\beta'J(1-q)} - \frac{\beta'Jq}{1-\beta'J(1-q)}}.
    \label{eq-app:log_det_approximation}
\end{align}
Using the approximation in \eqref{eq-app:log_det_approximation}, extremisation of \eqref{eq-app:RS_solution_Z} with respect to $H_a$ yields:
\begin{equation}
     m_a = \frac{1}{N}\sum_i \xi_i^a\int Dz\tanh \pr{ \beta'  \sum_{b\leq l} (H_b +J m_b) \xi_i^b +  \beta' J \sqrt{\alpha r} z }.
\end{equation}
Similarly, extremisation with respect to $r$ yields
\begin{align}
    \frac{ \beta'^2 J^2}{2} N\alpha (1-q) &=  N\beta' J \frac{1}{2} \sqrt{\frac{\alpha}{r}}  \int Dz\, z \tanh \pr{ \beta' \sum_{b\leq l} (H_b +J m_b) \xi_i^b+  \beta' J \sqrt{\alpha r} z },
\end{align}
which, by applying the partial integration, results in 
\begin{align}
    q &= \int Dz \tanh^2\pr{\beta' \pr{ \sum_{b\leq l} (H_b +J m_b) \xi_i^b + J \sqrt{\alpha r} z  }}.
\end{align}
We can observe that for $\alpha=0$, we recover previous results in \eqref{eq-app:mf-Hopfield}. In addition, we obtain Eqs.~\ref{eq:m_near_sat} and \ref{eq:q_near_sat} for one memory pattern ($l=1$) given by all positive unity values and $H_b=0$. Further, for $\gamma=0$, $\beta'=\beta$, and the solution corresponds to the Hopfield model near saturation \cite{amit1985storing,coolen2001statistical}.

Notice that in the limit $J\to 0$ and $\bm H=\bm 0$, we obtain
\begin{align}
    \beta' J^2\mu &=  \frac{1}{2} N J\alpha (1 +\beta' J (1-q^2))  - N\frac{J\alpha}{2}, 
    \\ r &=  q,
    \\ \beta' &= \frac{\beta}{1+\gamma N \beta  \frac{1}{2} \alpha \beta' J^2 (1-q^2) }.
\end{align}
For a scaled value of spin coupling strength, defined as $\tilde J$ such that $\alpha=1$ and $\gamma' = \gamma N \beta \frac{J^2}{\tilde J^2}$, the equations above 
recover the solution for the curved Sherrington-Kirkpatrick model in Eqs.~(\ref{eq-app:m_replica-symmetry-solution-SK}-\ref{eq-app:beta_replica-symmetry-solution-SK}).

\subsection{AT-instability line}

This section probes how the deformation of the statistics modifies the boundary below which we may no longer rely on replica symmetry. Following~\cite{coolen2001statistical} let us then consider small fluctuations $\eta_{uv}$ around the replica symmetric expressions for $q_{uv}$ and its conjugated pair.
\begin{equation}
    q_{uv} \mapsto q_{uv}^{\mathrm{RS}} + \eta_{uv} \coloneqq \delta_{uv} + q(1-\delta_{uv}) + \eta_{uv}
\end{equation}
with $\eta_{uv}=\eta_{vu}$, vanishing diagonal elements, and $\sum_u \eta_{uv}=0$. We are ultimately interested in the free energy difference,
\begin{equation}
\tfrac{1}{N}\Delta\varphi_\gamma \coloneqq \tfrac{1}{N}[\varphi_{\gamma}(m^{\mathrm{RS}},q_{uv},\hat{q}_{uv}) - \varphi_{\gamma}(m^{\mathrm{RS}},q_{uv}^{\mathrm{RS}},\hat{q}_{uv}^{\mathrm{RS}})].
\end{equation}
One should be mindful that $\beta'_u$ may be affected by fluctuations. The effective inverse temperature $\beta'_u$ depends on $\mu$, which is itself a function of both $q_{uv}$ and $\hat{q}_{uv}$. One can anticipate that $\hat{q}_{u v}$, and thereby $\beta'_u$, will be a polynomial in $\eta_{u v}$. The coefficients of the perturbative expansion of $\beta'_u$ are determined by replica-symmetric parameters, and hence its index structure follows from the properties of $\eta_{uv}$ rule out linear contributions. Without loss of generality, we have up to the second order,
\begin{equation}
    \beta' = \beta'_{0} + \beta'_{1}\sum_{v} \eta_{uv}^2 + \mathcal{O}(\bm \eta^3), 
\end{equation}
for some $\beta'_1$ to be determined and $\beta'_0$ being its RS-value, which only distinguishes between diagonal and off-diagonal components. Let us first recall that $\beta'_{0}$ at~\eqref{eq-app:saddle_node_sol} under the RS assumption becomes,
\begin{equation}
    \beta'_0 = \frac{\beta}{1+ \frac{1}{2}\gamma' ( \bar{m} + \alpha \beta'_0 J^2 (R - qr)- J\alpha)}
\end{equation}
with $\bar{m} = 2\sum_{b\leq l} \big(H_b m_b+ \frac{J}{2} (m_b)^2 \big)$ adopted for brevity. Solving for $\beta'_0$ from the expression above leads to,
\begin{equation}
    \beta'_{0} = \frac{2- J\alpha \gamma' + \gamma' \bar{m} \mp \sqrt{\left(2- J\alpha \gamma' + \gamma' \bar{m}\right)^2 + 8 \beta J^2 \alpha \gamma' (R-q r)}}{2\alpha \gamma' (q r- R)}.
\end{equation}
To resolve how its conjugate, $\hat{q}_{uv}$, transforms, we inspect the two-point functions $\ang{z_u z_v}_*$ upon small perturbations of the order parameter $\eta_{uv}$,
\begin{align}
    \ang{z_u z_v}_* \mapsto \frac{\ang{z_a z_b}_* + \frac{1}{2}\sum_{c,d}\ang{z_a z_b z_c z_d}_* \Lambda_{c d}}{1 + \frac{1}{2}\sum_{c,d}\ang{z_c z_d}_* \Lambda_{c d}} \simeq \ang{z_a z_b}_* + \frac{1}{2}\sum_{c,d}\Lambda_{c d}[\ang{z_a z_b z_c z_d}_* - \ang{z_a z_b}_*\ang{z_c z_d}_*]
\end{align}
with
\begin{equation}
    \Lambda_{cd} = \beta'_0 J \eta_{cd} + \beta'_1 J \sum_{s}(\eta_{cs}^2 + \eta_{ds}^2) [\delta_{cd} + q(1-\delta_{cd})].
\end{equation}
This implies that $\hat{\eta}_{cd}$, defined as the change of the two point function, and thereby $\hat{q}_{uv}$ ---conjugate to $q_{uv}$--- carries a dependence of second order in fluctuations parametrised by $\beta'_1$.
\begin{align}
    \hat{q}_{uv} &= -\iu N \alpha \sqrt{\beta'_u \beta'_v}J\ang{z_u z_v}_* \nonumber \\
    &\to -\iu N \alpha J \bigg( \beta'_0 + \frac{1}{2}\beta'_1 \sum_{s}[\eta_{us}^2 + \eta_{vs}^2] \bigg) [\ang{z_u z_v}_* + \hat{\eta}_{uv}] \nonumber \\
    &= \hat{q}^{\mathrm{RS}}_{uv} -\iu N \alpha J\bigg(\beta'_0 \hat{\eta}_{uv} + \frac{1}{2}\beta'_1 \sum_{s}\ang{z_u z_v}_* [\eta_{us}^2 + \eta_{vs}^2] \bigg).
\end{align}
Here $\beta'_{1}$ can be obtained from the expression for $\beta'_u$ at~\eqref{eq-app:saddle_node_sol} exploring $\sum_v \beta'_{v} q_{uv} r_{uv}$ under perturbations.
\begin{align}
    \sum_{v} q_{uv} \ang{z_u z_v}_* &\mapsto \sum_{v}(q_{uv}^{\mathrm{RS}} + \eta_{uv})(\ang{z_u z_v}_* + \hat{\eta}_{uv}) \nonumber \\
    &= \beta'_{0} J(R-rq) + \sum_v  (q_{uv}^{\mathrm{RS}}\hat{\eta}_{uv} + \ang{z_u z_v}_* \eta_{uv} + \eta_{uv} \hat{\eta}_{uv}) \nonumber \\
    &= \beta'_{0} J(R-rq) + \frac{1}{2}\beta'_1 \sum_{v,s,c,d}q_{uv}^{\mathrm{RS}}q_{cd}^{\mathrm{RS}}g_{uvcd}(\eta_{cs}^2 + \eta_{ds}^2)+ \frac{1}{2}\beta'_0 \sum_{v,c,d}g_{uvcd} \eta_{uv}\eta_{cd}, \label{eq-app:beta_pert_exp}
\end{align}
where the first term results from its RS-valued part and
\begin{equation}
    g_{abcd} = \ang{z_a z_b z_c z_d}_* - \ang{z_a z_b}_*\ang{z_c z_d}_* = \ang{z_a z_c}_*\ang{z_b z_d}_* + \ang{z_a z_d}_*\ang{z_b z_c}_* 
\end{equation}
has been adopted for brevity. The four-point function can be reduced via Wick Theorem to products of two-point functions. Linear terms in fluctuations coupled to RS-terms vanish with the sum as expected. It should be noted that unlike the flat case, the $\sum_s \hat{\eta}_{us} = 0$ property no longer holds due to quadratic terms in perturbations.\\

Let us now evaluate the sums at~\eqref{eq-app:beta_pert_exp},
\begin{align}
    \sum_{v,s,c,d}q_{uv}^{\mathrm{RS}}q_{cd}^{\mathrm{RS}}g_{uvcd}(\eta_{cs}^2 + \eta_{ds}^2) &= \sum_{v,s,c,d}q_{uv}^{\mathrm{RS}}q_{cd}^{\mathrm{RS}} (\ang{z_u z_c}_*\ang{z_v z_d}_* + \ang{z_u z_d}_*\ang{z_v z_c}_*) (\eta_{cs}^2 + \eta_{ds}^2) \nonumber \\
    &= (\beta'_0 J)^2 \sum_{v,s,c,d}q_{uv}^{\mathrm{RS}}q_{cd}^{\mathrm{RS}} (r_{uc}^{\mathrm{RS}}r_{vd}^{\mathrm{RS}} + r_{ud}^{\mathrm{RS}}r_{vc}^{\mathrm{RS}}) (\eta_{cs}^2 + \eta_{ds}^2)
\end{align}
from here, we break the sums into diagonal and off-diagonal contributions
\begin{align}
    &= 4 (\beta'_0 J)^2 \sum_{v,s,c}q_{uv}^{\mathrm{RS}} r_{uc}^{\mathrm{RS}}r_{vc}^{\mathrm{RS}} \eta_{cs}^2 + 2 q (\beta'_0 J)^2 \sum_{v,s,c \neq d}q_{uv}^{\mathrm{RS}} (r_{uc}^{\mathrm{RS}}r_{vd}^{\mathrm{RS}} + r_{ud}^{\mathrm{RS}}r_{vc}^{\mathrm{RS}}) \eta_{cs}^2 \nonumber \\
    &= 4 (\beta'_0 J)^2 \sum_{s,c} r_{uc}^{\mathrm{RS} 2} \eta_{cs}^2 + 4q (\beta'_0 J)^2 \sum_{v\neq u,s,c} r_{uc}^{\mathrm{RS}}r_{vc}^{\mathrm{RS}} \eta_{cs}^2
    + 4 q (\beta'_0 J)^2 \sum_{s,c \neq d} r_{ud}^{\mathrm{RS}}r_{uc}^{\mathrm{RS}} \eta_{cs}^2 \nonumber \\
    &\quad +2 q^2 (\beta'_0 J)^2\sum_{s,v\neq u,c \neq d} (r_{uc}^{\mathrm{RS}}r_{vd}^{\mathrm{RS}} + r_{ud}^{\mathrm{RS}}r_{vc}^{\mathrm{RS}}) \eta_{cs}^2.
\end{align}
Evaluation of $r_{ud}^{\mathrm{RS}} = R\delta_{ud} + r(1-\delta_{ud})$ yields a polynomial we will call for the moment $f(q,\beta'_0,r,R)$, and so the expression may be succinctly written as $f(q,\beta'_0,r,R)\sum_{s}\eta_{us}^2$. More importantly,
\begin{align}
    \sum_{v,c,d}g_{uvcd} \eta_{uv}\eta_{cd} &= (\beta'_0 J)^2 \sum_{v,c,d}(r_{uc}^{\mathrm{RS}}r_{vd}^{\mathrm{RS}} + r_{ud}^{\mathrm{RS}}r_{vc}^{\mathrm{RS}})\eta_{uv}\eta_{cd} \nonumber \\
    &= (\beta'_0 J)^2 \sum_{v\neq u,c\neq d}(r_{uc}^{\mathrm{RS}}r_{vd}^{\mathrm{RS}} + r_{ud}^{\mathrm{RS}}r_{vc}^{\mathrm{RS}})\eta_{uv}\eta_{cd} \nonumber \\
    &= 2(\beta'_0 J)^2 \sum_{v\neq u,c\neq d} r_{uc}^{\mathrm{RS}}r_{vd}^{\mathrm{RS}}\eta_{uv}\eta_{cd} = 2(\beta'_0 J)^2 \sum_{c\neq d} r_{uc}^{\mathrm{RS}} \bigg(\sum_{v\neq u} r_{vd}^{\mathrm{RS}}\eta_{uv}\bigg)\eta_{cd} = 0.
\end{align}
Now we can solve for $\beta'_1$, expanding the self-consistent equation,
\begin{align}
    \beta'_0 + \beta'_1 \sum_s \eta_{vs}^2 &\simeq \beta'_0 \bigg(1 - \frac{1}{4 \Gamma} \gamma' \alpha J \bigg( \beta'_1 \sum_{v,s,c,d}q_{uv}^{\mathrm{RS}}q_{cd}^{\mathrm{RS}}g_{uvcd}(\eta_{cs}^2 + \eta_{ds}^2) + \beta'_0 \sum_{v,c,d}g_{uvcd} \eta_{uv}\eta_{cd} \bigg) \bigg) \nonumber \\
    \beta'_1 \sum_s \eta_{vs}^2 &\simeq - \frac{1}{4 \Gamma} \gamma' \alpha J \beta'_1 f(q,\beta'_{0},r,R)\sum_s \eta_{us}^2,
\end{align}
where $\Gamma$ is defined as denominator of the expression for $\beta'$ at~\eqref{eq:beta_RS},
\begin{equation}
    \Gamma \coloneqq 1+ \frac{1}{2}\gamma' \pr{  J m^2 + \alpha  J ( \beta' (R - qr) - 1)}
\end{equation}
seeming to imply that $\beta'$ does not seem to be altered at the second order of perturbations, and perhaps the effects are only seen at higher orders. This greatly simplifies the analysis onwards; $\hat{q}_{uv}$ and $\hat{\eta}_{uv}$ are now first order in $\eta_{u v}$, the latter reduced to $\hat{\eta}_{uv} = \beta^{\prime 2}_0 (R-r)^2\eta_{uv}$. The same expression for $\gamma=0$ up to a scaled inverse temperature is recovered. Let us focus on the free energy difference,
\begin{align} \label{eq-app:qqhat_fluctuation}
      \frac{1}{N}\Delta\varphi_\gamma \ni \frac{1}{2 N}\sum_{u,v} (\iu \hat{q}_{uv}q_{uv} - \iu \hat{q}_{uv}^{\mathrm{RS}}q_{uv}^{\mathrm{RS}}) &= \frac{\iu}{2 N} \Tr\, [\hat{q}_{uv}^{\mathrm{RS}}\eta_{uv}] + \frac{1}{2}\alpha (\beta' J)^2 \Tr\, [\hat{\eta}_{uv}\hat{q}_{uv}^{\mathrm{RS}}+\hat{\eta}_{uv}\eta_{uv}] \nonumber \\
      &= \frac{1}{2}\alpha (\beta' J)^2 \Tr\, [\hat{\eta}_{uv}\eta_{uv}] .
\end{align}
The trace is understood over replica indices. Notice that despite the seemly different overall coefficient and sign~\cite{coolen2001statistical}, this is just an artifact of the convention on the introduction of the deltas; these expressions are equivalent. The diagonal terms are included to make up the trace vanish as constants at the free energy difference.\\

The determinants transform as,
\begin{equation}
    \frac{1}{N}\Delta\varphi_\gamma \ni \ln \frac{\abs{1 -\beta' (\bm q^{\mathrm{RS}}+\bm\eta)}}{\abs{1 -\beta' \bm q^{\mathrm{RS}}}} = -\frac{1}{2}\frac{\beta^{\prime 2}_0}{[1-\beta'_0(1-q)]^2}\Tr \bm \eta^2 + \mathcal{O}(\bm \eta^3).
\end{equation}
There is a contribution from the $L(x_i)$-function and the logarithm that results from the deformation. First the $L$-term contribution
\begin{equation}
    \frac{1}{N}\Delta\varphi_\gamma \ni \ln \frac{\sum_{\bm x} \exp{L(m^{\mathrm{RS}}, q_{uv}, \hat{q}_{uv},x_i)}}{\sum_{\bm x} \exp{L(m^{\mathrm{RS}},q_{uv}^{\mathrm{RS}}, \hat{q}_{uv}^{\mathrm{RS}},x_i)} }, 
\end{equation}
which basically amounts to, after expansion,
\begin{align}
     \ln \sum_{\bm x}\exp L(m^{\mathrm{RS}},q_{uv},\hat{q}_{uv},x_i) &\simeq \ln\prod_{i} \sum_{\bm x_i} \exp \Lambda^{\mathrm{RS}}(x_i) \bigg[1+ \alpha J^2 \sum_{u,v} x_{i}^u \hat{\eta}_{uv}x_{i}^v \sum_{u,v} \beta'_{0 u}\eta_{uv}\beta'_{1 v} \nonumber \\
     &\qquad \qquad \frac{1}{2}\alpha J^2\beta_{0}^{\prime 2} \sum_{u,v} x_{i}^u \hat{\eta}_{uv}x_{i}^v + \frac{1}{8}\alpha^2 (\beta'_0 J)^4\sum_{u,v} (x_{i}^u \hat{\eta}_{uv}x_{i}^v)^2 \bigg].
\end{align}
However, as we concluded previously, $\beta'_u$ does not have a second-order term.
\begin{equation}
    =\ln\prod_{i} \sum_{\bm x_i} \exp \Lambda^{\mathrm{RS}}(x_i) \bigg[1+ \frac{1}{2}\alpha (\beta'_0 J)^2 \sum_{u,v} x_{i}^u \hat{\eta}_{uv}x_{i}^v + \frac{1}{8}\alpha^2 (\beta'_0 J)^4\sum_{u,v} (x_{i}^u \hat{\eta}_{uv}x_{i}^v)^2 \bigg],
\end{equation}
where $\Lambda^{\mathrm{RS}}(x_i)$ has been defined as the argument of the exponential at~\eqref{eq_app:RS_Lfunction}. Once again, the trace can be recovered at the fluctuation terms can be recovered noticing that $\eta_{uu}=0$. The denominator eventually cancels off the contributions from $\Lambda^{\mathrm{RS}}(x_i)$, and we are left with the part from the squared brackets.
The first term can be recognised as the average defined at the saddle-node solution for $q_{ab}$~\eqref{eq-app:saddle_node_sol}. 
Finally, the logarithm that results from the deformation of the statistics and $\mu$,
\begin{equation}
    \frac{1}{N}\Delta\varphi_\gamma  \ni \frac{\beta}{\gamma'}\ln \left(\frac{\beta'_{\mathrm{RS}}}{\beta'}\right) -\beta' J \Delta \mu =0
\end{equation}
are up to second order in perturbations, invariant, and hence do not contribute to the free energy difference. Following the derivation of $\Delta \varphi_{\gamma}$ for $\gamma \to 0$ at~\cite{coolen2001statistical}, we may determine,
\begin{align}
    \frac{1}{N}\Delta \varphi_\gamma = \frac{1}{\beta'_0
    n}\Bigg(&-\frac{1}{4}
    \frac{\alpha \beta^{\prime 2}_0}{\left[1-\beta'_0(1- q)\right]_+^2}
    \Tr \bm \eta^2
    +\frac{1}{2}\alpha\beta^{\prime 4}_0 (R-r)^2 \Tr \bm \eta^2 \nonumber \\
    & -\frac{1}{8}\alpha^2\beta^{\prime 8}_0(R- r)^4
\sum_{a,b,c,d}\eta_{a b}\eta_{cd}G_{abcd}\Bigg),
\end{align}
where
\begin{align}
    G_{abcd} &=
    \delta_{ac}\delta_{bd} + \delta_{ad}\delta_{bc} + G_4 (1-\delta_{ac})(1-\delta_{bd})(1-\delta_{ad})(1-
    \delta_{bc}) \nonumber \\
    &\quad\;+ G_2 (\delta_{ac}(1-\delta_{bd})+ \delta_{bd}(1-\delta_{ac})+ \delta_{ad}(1-\delta_{bc})+ \delta_{bc}(1-\delta_{ad})),
\end{align}
with
\begin{equation}
    G_\ell = \frac{1}{N}\sum_{i}\frac{\int Dz \tanh^\ell \beta'_0 [\sum_{b\leq l} (H_b + J m_b)\xi^{b}_i + z \sqrt{\alpha r} ] \cosh^n \beta'_0 [\sum_{b\leq l} (H_b + J m_b)\xi^{b}_i + z \sqrt{\alpha r} ]}{\int Dz \cosh^n \beta'_0 [\sum_{b\leq l} (H_b + J m_b)\xi^{b}_i + z \sqrt{\alpha r} ]}.
\end{equation}
leading to a condition
\begin{equation}
    (1+ \beta'_0 (1-q))^2 > \alpha \beta^{\prime 2}_0 \frac{1}{N}\sum_i \xi_i^a\int Dz \cosh^{-4} \beta'_0 \bigg( \sum_{b\leq l} (H_b + J m_b)\xi^{b}_i + J \sqrt{\alpha r} z \bigg)
\end{equation}
equivalent to that of the flat model (see~\cite{coolen2001statistical}, equation (121)) with a rescaled inverse temperature.

\newpage
\section{Curved Sherrington-Kirkpatrick model}
\label{app:SK}

We start with a simple case in which the system is encoding one pattern on a background of zero-average Gaussian weights. This can be represented by $J_{ij} = J_0/N \xi_i\xi_j + J/\sqrt{N}  z_{ij}$, with $z_{ij}$ random coupling values distributed as $\mathcal{N}(0,1)$. Assuming the content of the $\brpos{\,}$ operator is positive, we want to compute the configurational average
\begin{align}
    \llangle \varphi_\gamma \rrangle&= \int D\bm z  \ln \sum_{\bm x} \Bigg(1+\gamma \beta \Bigg(\frac{J_0}{N} \sum_{i<j} x_i \xi_i \xi_j x_j + \frac{J}{\sqrt N}\sum_{i<j} z_{ij} x_i x_j \Bigg)\Bigg)^{1/\gamma}
    \\\nonumber &= \int D\bm z \ln \sum_{\bm x} \Bigg(1+\gamma \beta \Bigg(  \frac{NJ_0}{2} \bigg(\frac{1}{N}\sum_{i} x_i \xi_i \bigg)^2 - \frac{J_0}{2} + \frac{J}{\sqrt N}\sum_{i<j}  z_{ij}  x_i x_j\Bigg)\Bigg)^{1/\gamma},
\end{align}
where we define $D\bm z = \prod_{i<j} d z_{ij} \frac{1}{\sqrt{2\pi}} \exp\pr{-\frac{1}{2}z_{ij}^2}$.
 
Defining $\varphi_\gamma=\ln Z$, we can apply the replica method
\begin{equation}
    \llangle \varphi_\gamma \rrangle =\llangle \ln Z \rrangle = \lim_{n\to 0} \frac{1}{n}(\llangle Z^n \rrangle - 1) \qquad\mathrm{or, equivalently,} \qquad \llangle \ln Z \rrangle = \lim_{n\to 0} \frac{1}{n}\ln\llangle Z^n \rrangle,
\end{equation}
with 
\begin{align}
    \llangle Z^n \rrangle&= \int D\bm z  \sum_{\bm x} \exp\pr{\sum_u\frac{1}{\gamma}\ln \brpos{1+\gamma \beta \Bigg( \frac{NJ_0}{2} \bigg(\frac{1}{N}\sum_{i} x_i^u \xi_i \bigg)^2 - \frac{J_0}{2} + \frac{J}{\sqrt{N}}\sum_{i<j}  z_{ij} x_i^u  x_j^u \Bigg)}} 
     \\ &= \frac{1}{(2\pi)^{2n}} \int d\bm z p(\bm z)  \sum_{\bm x}  d\bm m d\bm{\hat m}d\bm\mu d\bm{\hat \mu}\exp\vast( \frac{1}{\gamma}\sum_{u} \ln\pr{1+\gamma \beta \pr{ \frac{NJ_0}{2} m_u^2 - \frac{J_0}{2} + J \mu_u}} \nonumber
    \\&\quad\, - \sum_u  \iu\hat m_u \bigg(m_u - \frac{1}{N}\sum_{i} \xi_i x_i^u \bigg)-  \sum_{u} \iu\hat \mu_{u} \Bigg(\mu_{u}- \frac{1}{\sqrt{N}}\sum_{i< j}z_{ij} x_i^u x_j^u \Bigg)\vast).
\end{align}
Recalling that the $\bm z$ couplings are distributed by a centred Gaussian $\mathcal{N}(0,1)$, we can carry out explicit integration of $z_{ij}$. Noting that $\int D\bm z\, e^{\bm \lambda^\top \bm z}= e^{\bm \lambda^\top \bm \lambda /2}$, the above may be rewritten as, 
\begin{align}
    &=  \frac{1}{(2\pi)^{2n}} \int  d\bm m d\bm{\hat m}d\bm\mu d\bm{\hat \mu}\sum_{\bm x} \exp\vast( \frac{1}{\gamma}\sum_{u} \ln\pr{1+\gamma \beta \pr{ \frac{NJ_0}{2} m_u^2 - \frac{J_0}{2} + J \mu_u}} \nonumber
    \\&\quad\, - \sum_u \iu\hat m_u \bigg( m_u - \frac{1}{N}\sum_{i}\xi_i  x_i^u \bigg)-  \sum_{u} \iu\hat \mu_{u} \mu_{u}+ \frac{1}{4N} \sum_{i\neq j} \bigg(\sum_u \iu\hat\mu_u x_i^u x_j^u \bigg)^2\vast).
\end{align}
The last term in the exponential can be expressed as,
\begin{equation}
    N\frac{1}{2}\sum_{u<v} \iu\hat\mu_u  \iu\hat\mu_v  \Bigg(\bigg(\frac{1}{N}\sum_i x_i^u  x_i^v\bigg)^2-\frac{1}{N}\Bigg) - N\frac{1}{4}\sum_{u} \pr{\iu\hat\mu_u }^2 \pr{1-\frac{1}{N}}.
\end{equation}
Furthermore, introducing conjugate pair fields for the average of $x_{i}^u x_{i}^v$, $\{\bm q, \bm{\hat{q}}\}$ we have,
\begin{align}
    & (2\pi)^{-2n - n(n-1)}\int  d\bm m d\bm{\hat m}  d\bm\mu d\bm{\hat \mu} d\bm q d\bm{\hat q} \nonumber \\
    & \sum_{\bm x} \exp\vast( \frac{1}{\gamma}\sum_{u} \ln\pr{1+\gamma \beta \pr{ \frac{NJ_0}{2} m_u^2 - \frac{J_0}{2} + J \mu_u}} - N\frac{1}{4}\sum_{u} \pr{\iu\hat\mu_u }^2 \pr{1-\frac{1}{N}} - \sum_{u} \iu\hat \mu_{u} \mu_{u} \nonumber
    \\&  - \sum_u  \iu\hat m_u \pr{m_u - \frac{1}{N}\sum_{i} \xi_i x_i^u} - \sum_{u<v}\iu\hat q_{uv}\pr{ q_{uv} - \frac{1}{N}\sum_i x_i^u x_i^v}  +  N\frac{1}{2}\sum_{u<v} \iu\hat\mu_u  \iu\hat\mu_v \pr{q_{uv}^2 - \frac{1}{N}} \vast).
\end{align}
Now we can evaluate the integrals by steepest descent
\begin{align}
    &=  \exp\vast\{ \frac{1}{\gamma}\sum_{u} \ln\pr{1+\gamma \beta \pr{ \frac{NJ_0}{2} m_u^2 - \frac{J_0}{2} + J \mu_u}}  - N\frac{1}{4}\sum_{u} \pr{\iu\hat\mu_u }^2 \pr{1-\frac{1}{N}} \nonumber
    \\&  - \sum_u  \iu\hat m_u m_u   - \sum_{u} \iu\hat \mu_{u} \mu_{u}  - \sum_{u<v} \iu\hat q_{uv} q_{uv}  + \frac{1}{2}N\sum_{u<v}\iu\hat\mu_u  \iu\hat\mu_v  \pr{ q_{uv}^2 - \frac{1}{N}}   + \ln\sum_{\bm x} \exp L  \vast\}. \label{eq-app:Zn_steepest}
\end{align}
The overall $2\pi$ factor has been left out as we are ultimately concerned with $n\to 0$. With $L$ corresponding to the $x_i$-dependant part in the argument of the exponential~\eqref{eq-app:L_function}. Here, \eqref{eq-app:Zn_steepest} is understood at the saddle-node solution, which, ignoring the $\mathcal{O}(\frac{1}{N})$ terms, corresponds to,
\begin{align} 
\label{eq-app:saddle-sol}
\begin{aligned}
    \iu\hat m_u &= \beta'_u N J_0 m_u,
    \\ \iu\hat\mu_{u} &= \beta'_u J ,
    \\  m_u &= \ang{x_i^u } ,
    \\  q_{uv} &= \ang{x_i^u x_i^v} ,
    \\  i \hat q_{uv} &= N\iu\hat\mu_{u}\iu\hat\mu_{v} q_{uv} = N \pr{ \beta'_u J }^2 q_{uv},
    \\\mu_{u} &=   N \frac{1}{2} \sum_{v}\iu\hat\mu_{v} q_{uv}^2  =  N J \frac{1}{2} \sum_{v} \beta'_v q_{uv}^2, 
    \\ \beta_u' &= \frac{\beta}{1+\gamma \beta \pr{ N\frac{1}{2}J_0 m_u^2+ \mu_{u} }}  = \frac{\beta}{1+\gamma \beta N \frac{1}{2}\pr{J_0 m_u^2 +  J^2\sum_{v}\beta'_v q_{uv}^2 }}. 
\end{aligned}
\end{align}
Assuming $q_{uu}=1$, we can rewrite~\eqref{eq-app:Zn_steepest} by evaluating at~\eqref{eq-app:saddle-sol}. As we are contemplating the $n\to 0$ limit, $N$ is taken large but kept at a fixed value, resulting in
\begin{align}
    \llangle \varphi_\gamma \rrangle  &= \frac{1}{n}\frac{1}{\gamma}\sum_{u} \ln\pr{1+\gamma \beta \pr{ \frac{NJ_0}{2} m_u^2 - \frac{J_0}{2} + N J^2 \frac{1}{2} \sum_{v} \beta'_v q_{uv}^2 }} \nonumber
    \\&\quad -\frac{1}{n} \sum_u \beta'_u N J_0 m_u^2 - N \frac{3}{4n}\sum_{u,v} \beta'_u   \beta'_v J^2 q_{uv}^2  + \frac{1}{n}\ln\sum_{\bm x} \exp L.
\end{align}

In the limit $\gamma\to 0$, we recover the replica free-energy of the SK model
\begin{equation}
    \lim_{\gamma\to 0}\llangle \varphi_\gamma \rrangle =- \beta N \frac{J_0}{2n} \sum_u    m_u^2   - N \frac{\beta^2 J^2 }{4n}\sum_{u,v} q_{uv}^2  + \frac{1}{n}\ln\sum_{\bm x} \exp L. 
\end{equation}

\subsection{Replica symmetry}
The assumption of replica symmetry implies homogeneous couplings among replicas $q_{uv}= \delta_{uv} + q(1-\delta_{uv})$. Also we will consider $m_u = m$ for the mean field,
\begin{align}
    \llangle \varphi_\gamma \rrangle = & \frac{1}{\gamma} \ln\brpos{1+\gamma \beta \pr{ \frac{NJ_0}{2} m^2 - \frac{J_0}{2} + N J^2 \frac{1}{2}\beta' (1- q^2) }} \nonumber  
    \\& - \beta' N J_0 m^2 - N \frac{3}{4} \beta'^2 J^2 (1-q^2 ) + \frac{1}{n}\ln\sum_{\bm x} \exp L 
\end{align}
with $L$ carrying the $\bm x$ dependence. We can further simplify evaluating the sum,
\begin{align}
    \ln\sum_{\bm x} \exp L &= \ln\sum_{\bm x}  \exp\pr{ n \beta' J_0 m \sum_{i,u} \xi_i  x_i^u + (\beta' J)^2 q \sum_{i,u<v} x_i^u  x_i^v } \nonumber
    \\ &= \ln\prod_i \sum_{\bm x_i} \exp\pr{ \sum_{u}\beta' J_0 m \xi_i x_i^u  + \frac{1}{2}   \pr{ \beta' J \sqrt{q} \sum_{u} x_i^u }^2 - \frac{1}{2} n \pr{\beta' J}^2 q } \nonumber
    \\ &= \ln\prod_i \int Dz \sum_{\bm x_i} \exp\pr{ \sum_{u} \beta' J_0 m  x_i^u +  \beta' J \sqrt{q} z \sum_{u} x_i^u  - \frac{1}{2} n \pr{\beta' J}^2  q } \nonumber
    \\ &= N \ln \int Dz \exp\pr{ n \ln\pr{2 \cosh\pr{ \beta'  J_0 m +  \beta' J \sqrt{q} z  }}}- \frac{1}{2} nN \pr{\beta' J}^2 q \nonumber
    \\ &\approx Nn \int Dz \ln\pr{2 \cosh\pr{ \beta' J_0 m +  \beta' J \sqrt{q} z }}- \frac{1}{2} nN \pr{\beta' J}^2 q. 
\end{align}
We thus obtain
\begin{align}
    \frac{1}{N}\llangle \varphi\rrangle = &  \frac{1}{N\gamma} \ln\pr{1+N\gamma \beta \pr{ \frac{J_0}{2} m^2  +   J^2 \frac{1}{2}\beta' (1- q^2) }} \nonumber
    \\&  -    \beta'  J_0 m^2   -  \frac{3}{4} \beta'^2 J^2 (1-q^2 ) +   \int Dz \ln\pr{2 \cosh\pr{ \beta' J_0 m +  \beta' J \sqrt{q} z }}- \frac{1}{2}  \pr{\beta' J}^2 q. 
\end{align}
Extremisation with respect to $m$ and $q$ yields:
\begin{align}
    \beta' J_0 m &= \beta' J_0  \int Dz\tanh \pr{ \beta' J_0 m +  \beta' J \sqrt{q} z },
    \\    \frac{ \beta'^2 J^2}{2} (1-q) &= \beta' J \frac{1}{2\sqrt{q}}  \int Dz \tanh \pr{ \beta' J_0 m +  \beta' J^2\sqrt{q} z }  z,
\end{align}
leading to the solution, for $\gamma' = N \beta \gamma$:
\begin{align}
    m &= \int Dz \tanh\pr{\beta' (J_0 m  +  J \sqrt{q} z)},
        \label{eq-app:m_replica-symmetry-solution-SK}
    \\q &= \int Dz \tanh^2\pr{\beta' ( J_0 m  + J \sqrt{q} z  )},
    \label{eq-app:q_replica-symmetry-solution-SK}
    \\   \beta' &= \frac{\beta}{1+\gamma' \pr{ \frac{1}{2}J_0 m^2 +    \frac{1}{2}\beta' J^2(1- q^2) }}. 
    \label{eq-app:beta_replica-symmetry-solution-SK}
\end{align}

\subsubsection{Critical point}

The solution for  $J_0=0,J=1$ at $\gamma'=0$ has the form
\begin{equation}
    q = \int Dz \tanh^2\pr{\beta J\sqrt{q} z  }.
        \label{eq-app:q_replica-symmetry-flat-SK}
\end{equation}
Using a change of variables $\rho = \sqrt{q}$, we can expand around  $\rho\to 0$ ($\tanh^2 x = ( x - \frac{1}{3}x^2 + \cdots)^2 = x^2 - \frac{2}{3}x^4 + \cdots$ for small $x$). Noting that $\int Dz \, z^2=1$ and $\int Dz \, z^4=3$, we obtain
\begin{align}
    q &= \int Dz \tanh^2\pr{\beta J  \rho z  } = (\beta J)^{2} \rho^2 - 2 (\beta J)^{4} \rho^4 + \mathcal{O}(\rho^6)
    \nonumber \\&= (\beta J)^{2} q - 2 (\beta J)^{4} q^2+ \mathcal{O}(q^3).
\end{align}
This yields a trivial solution $q=0$, and a solution
\begin{equation}
    q= \frac{1}{2(\beta J)^4} ( (\beta J)^2 -1)
\end{equation}
for $\beta J >1$, with a slope of
\begin{equation}
    \frac{\partial q}{\partial \beta} =  \frac{1}{ \beta^5 J^4}(2-(\beta J)^2),
\end{equation}
which is equal to $1$ near the critical point, $\beta J=1$.

For $\gamma'\neq 0$, we can recover the critical solution by a change of variables $\beta\to\beta'$. For $m=0$, the solution of \eqref{eq-app:q_replica-symmetry-solution-SK} for an arbitrary $\beta'$ is the same as the solution of \eqref{eq-app:q_replica-symmetry-flat-SK} for $\beta=\beta'$.
For each pair of $\beta',q$ solving \eqref{eq-app:q_replica-symmetry-solution-SK}, we can recover the corresponding inverse temperature from \eqref{eq-app:beta_replica-symmetry-solution-SK} as $\beta = \beta'(1+\frac{1}{2} \gamma' \beta'J^2(1-q^2))$. 

Following the previous argument, we can show
\begin{equation}
    q= \frac{1}{2(\beta'J)^4} ( (\beta'J)^2 -1),
\end{equation}
resulting in that, at the critical $\beta$, we must obtain $\beta'J=1$ and $q=0$. Then we have
\begin{equation}
    \beta' +  \beta'^2 J^2 \gamma'  \frac{1}{2} = \beta,
\end{equation}
which, for $\beta'J=1$, yields the critical inverse temperature
\begin{equation}
    \beta_c= J^{-1}+\frac{1}{2}\gamma'.
\end{equation}

The derivative of $\beta'$ yields
\begin{align}
    \frac{d\beta'}{d\beta} (1+\beta'\gamma' J^2)&=1
    \\ \frac{d\beta'}{d\beta} &=\frac{1}{1+\beta'\gamma' J^2},
\end{align}
resulting in a slope of
\begin{equation}
    \frac{\partial q}{\partial \beta} = \frac{\partial q}{\partial \beta'}\frac{\partial \beta'}{\partial \beta}=  \frac{1}{ \beta'^5 J^4}\frac{2- (\beta'J)^2}{1+\beta'\gamma' J^2},
\end{equation}
which, for $\beta'=J^{-2}$, diverges at $\gamma'=-1$, resulting in a second-order phase transition.

\newpage
\section{Glauber dynamics of dense associative memories}
\label{app:dense}

The energy of the dense associative memories (a.k.a., modern Hopfield networks) with a state $\bm x=(x_1,\dots,x_n)\in\{-1,1\}^n$ takes the following functional form:
\begin{equation}
    \mathcal{F} = -\sum_{a} F\Big(\sum_i \xi_i^a x_i\Big),
\end{equation}
where $F(\cdot)$ is a non-linear function. The original formulation of dense associative memories used the rectified polynomial function $F(z)=z^p \cdot \Theta(z)$ with $\Theta(z)$ being the Heaviside step function~\cite{krotov2016dense}, and other authors have used the exponential function $F(z)=e^{z}$~\cite{demircigil2017model}. 

The deterministic update rule of the dense associative memories can be written using the following conditional probability as follows:
\begin{equation}
	p(x_k | \bm{x}_{\backslash k}) 
    = \Theta\big(\Delta \mathcal{F}(\bm x)\big),
    \label{eq-app:activation_function_modern}
\end{equation}
where $\Delta \mathcal{F}(\bm{x}) = \mathcal{F}(-x_k, \bm{x}_{\backslash k}) - \mathcal{F}(x_k, \bm{x}_{\backslash k})$. The energy difference $\Delta\mathcal{F}$ can be expressed as
\begin{align}
	\Delta \mathcal{F}(\bm{x})
    &= \sum_a F\Big(\sum_i \xi_i^a x_i\Big) - \sum_a F\Big(- \xi_k^a x_k + \sum_{j \neq k} \xi_j^a x_j\Big) \nonumber\\
    &=  \sum_a \bigg( F\Big(\sum_i \xi_i^a x_i\Big) -  F\Big(- 2 \xi_k^a x_k + \sum_{i} \xi_i^a x_i\Big) \bigg).
\end{align}
Thus, using the shorthand notation $\Delta \epsilon_k^a \coloneqq 2 \xi_k^a x_k$ (corresponding to the correlation between the $k$-th element of the $a$-th memory and the state) and 
\begin{equation}
    \Delta F_k^a \coloneqq F\Big(\sum_i \xi_i^a x_i\Big) -  F\Big(- \Delta \epsilon_k^a + \sum_{i} \xi_i^a x_i\Big), 
\end{equation}
the input to the threshold activation function in Eq.~\eqref{eq-app:activation_function_modern} can be expressed as
\begin{align}
	\Delta \mathcal{F}(\bm{x})
    = \sum_a  \frac{\Delta F_k^a}{\Delta \epsilon_k^a} \Delta \epsilon_k^a 
    = \sum_a w_{k}^{a} \Delta \epsilon_k^a, 
    \quad\text{where}\quad
    w_{k}^{a} = \frac{\Delta F_k^a}{\Delta \epsilon_k^a}. 
\end{align}
Above, we are assuming that $\xi_i^a \neq 0$ so that $\Delta \epsilon_k^a \neq0$ for simplicity. This implies that, in the dense associative memories, each neuron has distinct effective weights for each memory: the $k$-th neuron receives an input $\sum_a w_{k}^{a} \Delta \epsilon_k^a$, where $w_{k}^{a}$ weight $\Delta \epsilon_k^a$, which measures the matching of the state $x_k$ with the memory $\xi_k^a$. We can also regard the process as the gain modulation of the original weight $\xi_k^a$ attached to the input $x_k$ by $w_{k}^{a}$. 

Let us now showcase that this effective weight $w_{k}^{a}$ is an increasing function of $\sum_i \xi_i^a x_i$ for $F(z)=z^p \cdot \Theta(z)$ and also for $F(z)=e^z$, and hence it supports the accelerated retrieval of a selected memory. 

Let us first consider the case of $F(z)=z^p \cdot \Theta(z)$ with any integer $p \geq 2$. If $\Delta z > 0$, we have
\begin{equation}
    \Delta F(z) = F(z) - F(z- \Delta z) = 
    \begin{cases}
        0  \,\,\, & z < 0 \\
        z^p  & 0 \leq z < \Delta z \\
        z^p - (z-\Delta z)^p & z \geq \Delta z
    \end{cases}
\end{equation}
This function is non-negative for all $z$. It is also an increasing function of $z$ (for $z \geq \Delta z$, $\partial_z \Delta F(z) = p(z^{p-1} - (z-\Delta z)^{p-1}) > 0 $). Similarly, if $\Delta z < 0$, we have
\begin{equation}
    \Delta F(z) =
    \begin{cases}
        0  \,\,\, & z < \Delta z \\
        -(z-\Delta z)^p  & \Delta z \leq z < 0 \\
        z^p - (z-\Delta z)^p  & z \geq 0 
    \end{cases}
\end{equation}
It is a non-positive, decreasing function of $z$. Hence, $\frac{\Delta F(z)}{\Delta z}$ is a non-negative, increasing function of $z$. Namely, with $\Delta F(z) = \Delta F_k^a(z)$, $z=\sum_i \xi_i^a x_i$ and $\Delta z = \Delta \epsilon_k^a$, $w_{k}^{a} = \frac{\Delta F_k^a}{\Delta \epsilon_k^a}$ is a non-negative, increasing function of $\sum_i \xi_i^a x_i$. 

Let us now consider the case of $F(z)=e^z$. In this case we have $\Delta F(z)=e^{z} - e^{z-\Delta z}$, which is positive if $\Delta z > 0$ and negative if $\Delta z < 0$. The derivative is $\partial_z \Delta F(z)=e^z - e^{z-\Delta z}$, which is positive if $\Delta z>0$ and negative if $\Delta z<0$. This guarantees that $\frac{\Delta F_k^a}{\Delta \epsilon_k^a}$ is a positive, increasing function of $\sum_i \xi_i^a x_i$.

The proof can be extended to a differentiable function $F(z)$ if it is increasing $F'(z)>0$, and convex $F''(z) > 0$. Let $\Delta z > 0$. By the fundamental theorem of calculus, we have
\begin{align}
\frac{\Delta F(z)}{\Delta z}
   =\frac{1}{\Delta z}\int_{z-\Delta z}^{z} F'(t)\,\mathrm dt.
\end{align}
Because $F(z)$ is convex, $F'(z)$ is increasing, so the integral average on the right is an increasing function of $z$. Positivity follows from $F'(z)>0$. Similarly for $\Delta z < 0$
\begin{align}
\frac{\Delta F(z)}{\Delta z}
   =\frac{1}{\Delta z}\int_{z-\Delta z}^{z} F'(t)\,\mathrm dt
   =\frac{1}{-|\Delta z|}\int_{z+|\Delta z|}^{z} F'(t)\,\mathrm dt 
    =\frac{1}{|\Delta z|}\int_{z}^{z+|\Delta z|} F'(t)\,\mathrm dt 
\end{align}
to which the same argument applies. These prove $\frac{\Delta F(z)}{\Delta z}$ is a positive, increasing function of $z$. We note that the proof can be further extended to non-differentiable convex functions, too.

Therefore, in these systems, as $\sum_i \xi_i^a x_i$ increases (i.e., as the pattern $\bm \xi^a=(\xi_1^a,\ldots,\xi_N^a)$ is retrieved), the effective weights related to $a$ (i.e., $w_{k}^{a}$, $k=1,\ldots,N$) increase. This accelerates the alignment of $x_k$ with $\xi_k^{a}$, ensuring positive feedback. Additionally, retrieval of $\bm \xi^a$ reduces $\sum_i \xi_i^b x_i$ for orthogonal patterns $\bm \xi^b$, lowering their effective weights, thereby suppressing their recall and minimizing interference. This competitive mechanism highlights the superior capacity of these models compared to curved neural networks with uniform temperature scaling. Unlike the effective inverse temperature in the curved networks that depends only on the system's state or energy, the effective weight in updating $k$-th neuron in the dense associative memories additionally depends on the neuron's state $x_k$, thus no longer represents a global modulation of neurons.

\end{document}